\documentclass[12pt,preprint]{aastex}
\usepackage{xcolor}
\begin{document}
\title{Cataclysmic Variables observed during K2 Campaigns 0 and 1}
\author{Zhibin Dai\altaffilmark{1}}
\affil{Yunnan Observatories, Chinese Academy of Sciences.}
\affil{P. O. Box 110, 650011, Kunming, P. R. China.}
\email{zhibin\_dai@ynao.ac.cn}
\author{Paula Szkody}
\affil{University of Washington, Seattle, WA, 98195, USA.}
\author{Peter M. Garnavich and Mark Kennedy \altaffilmark{2}}
\affil{University of Notre Dame, Notre Dame, IN, 46556, USA.}
\altaffiltext{1}{ Key Laboratory for the Structure and Evolution of Celestial Objects, Chinese Academy of Sciences, P. R. China.}
\altaffiltext{2}{Department of Physics, University College Cork, Cork, Ireland.}

\begin{abstract}
There are 15 cataclysmic variables (CVs) observed in the first two campaigns of the K2 mission. In this paper, the eight CVs showing distinct features are analyzed in detail. Among these eight, modulations during quiescence are evident at the known orbital periods in the SU\,UMa stars QZ\,Vir and RZ\,Leo, and at our newly determined orbital periods of 1RXS\,J0632+2536 and WD\,1144+011. The periodogram analysis for the quiescent light curve of QZ\,Vir reveals multi-period modulations and the coexistence of orbital and superhump periods. The phased orbital light curves for the other 3 CVs in quiescence display wide (about half cycle) and shallow ($<$ 0.5\,mag) eclipse features. Besides these modulations, their quiescent light curves reveal several transient events: a sudden decrease of system light in 1RXS\,J0632+2536, a low level flare-like event in QZ\,Vir, a short brightening event in RZ\,Leo and a temporary disappearance of the orbital modulation in WD\,1144+011. The two known dwarf novae UV\,Gem and TW\,Vir and the CVs USNO-B1.01144-00115322 and CSS130516:111236.7+002807 show outbursts, including 1 complete and 3 incomplete normal outbursts and 2 complete superoutbursts. An incomplete but typical normal outburst confirms the dwarf nova identification of the USNO-B1.01144-00115322. The one complete normal outburst in UV\,Gem possibly provides the orbital period, since its modulations are shorter than the previously observed superhump period. The superoutbursts of TW\,Vir and CSS130516:111236.7+002807, along with their corresponding superhump periods, indicate that both objects are SU\,UMa stars. The derived superhump period of CSS130516:111236.7+002807 is 1.44\,hr, implying that this new SU\,UMa star is close to the period minimum.
\end{abstract}

\section{Introduction}

Cataclysmic variables (CVs) are interacting close binary systems consisting of a white dwarf primary and a Roche-lobe filling low-mass companion in a tight orbit  with a period of $\sim$\ 80\,min\,-\,10\,hr.  Some subsets of CVs with weak magnetic fields of the white dwarfs (B\,$<$\,10$^{6}$\,G), are classified as non-magnetic CVs including novalikes (NL), dwarf novae (DN), classic novae (CN) and recurrent novae (RN). In these systems, the material transferred from the secondary spreads into an accretion disk surrounding the primary due to the viscous processes between the adjacent accretion annulus (e.g. friction and shear). The viscosity within the accretion disk gives rise to the outward angular momentum migration and inward material flow through the disk releasing gravitational potential energy \citep[cf.][]{can88,war03,fra02}. The disk is typically the brightest component in a non-magnetic CV system and dominates most of the observed features, such as quasi-periodic DN outbursts of $\sim$\ 4\,-\,9\,mag and the nearly constant system light of the NL subtype. Although there are some discussions about the relationship between the accretion disk and a nova outburst caused by a thermonuclear runaway \citep[e.g.][]{ret98,lip02}, the accretion disk in a DN system is a dominant light source in optical, UV and X-ray bands and plays a key role in DN normal outbursts and superoutbursts (outbursts in short orbital period systems which are brighter and last longer than normal outbursts). Comprehensive reviews of DN outbursts in disk systems are given in \citet{osa96} and \citet{pat05}. 

During the four-year NASA Kepler mission from 2009 March to 2013 May \citep{bor10,haa10}, 15 of the 27 CVs located in the Kepler field were monitored successfully with results currently published for 8 \citep[e.g.][]{how13,sca13a,sca13b}. Since the orbital periods of most CVs are $\sim$\,1.5\,-\,4\,hrs, the current all-sky surveys (e.g. CRTS, ASAS and PanSTARRS) with the cadence of $\sim$\ 1\,-\,2\,obs/24 hrs at best can only provide information on outbursts but little on orbital variability during quiescence and outburst. Therefore, Kepler holds an enormous potential for stringent tests of accretion disk dynamics and precession at quiescence and outburst thanks to the unprecedented light curves with nearly continuous photometric coverage \citep[e.g.][]{woo11,can12,sma13,osa13,osa14}. Despite the end of the Kepler prime mission due to the loss of two of the spacecraft's reaction wheels, the K2 mission \citep{how14}, as a new ecliptic-plane mission with continuous high-precision data ($\sim$\ 50\,ppm from 6.5-hrs S/N for a 12th\,mag G star, close to the sensitivity of the primary mission) over an 80 day period for each campaign field, continues Kepler's ground-breaking discoveries on CVs.  In particular, the short-cadence observations in the K2 mission extend the studies of sporadic variability and the changes in the disk and stream impact area (i.e., hot spot) during normal outbursts and superoutbursts. 

In this paper, we report on the 15 CVs, of which 6 and 9 are in the K2 Campaign 0 (K2-C0) and Campaign 1 (K2-C1) fields, respectively.  These targets include 8 known DN, one possible intermediate polar (IP), one new DN candidate, one peculiar CV and 4 unknown type CVs from the Catalina Real-time Transient Survey (CRTS).

\section{K2 Observations and Data Reductions}

The RK catalogue (Version 7.21 released by \citet{rit03}), the Downes web catalogue (i.e., cvcat), the CRTS database \citep{dra09}, the SDSS database \citep{szk11} and SIMBAD were used to search all known CVs which likely fall within the fields of view of K2-C0 and K2-C1 by using the code K2fov developed by the Kepler team. There are 12 and 10 known CVs found in our proposals of C0 and C1, respectively, with 6 targets in C0 and 9 in C1 actually observed. Table 1 lists these 15 CVs. Although each campaign of the K2 mission is originally designed for a period of over 80 days, the available data provided by C0 only cover $\sim$\ 35\,days on average, since the Kepler spacecraft was not in fine point during the beginning part of C0, which was just an engineering test operated from March 12 to May 27, 2014. The following C1 as the first full length observing campaign from May 30 to August 20, 2014, provided long cadence (LC; 30\,min) observations for all objects as well as short cadence (SC; 1\,min) observations for the three DN: QZ Vir, TW Vir and RZ Leo. There is a 3-day data gap in C1 between BJD\,2456848 and BJD\,2456851.

\subsection{Light Curve Extractions}

The Mikulski Archive for Space Telescopes (MAST) provides Target Pixel Files (TPF) stored under the sole Ecliptic Plane Input Catalogue (EPIC) identification number (see Table 1). All 15 CV light curves were extracted from the original TPFs in C0 (data release 2) and C1 (data release 3) by using the program PyKE, which was developed by the Guest Observer Office \citep{sti12}. The four pipeline tasks of PyKE: kepmask, kepextract, kepdraw and kepconvert were used to reduce the TPFs. 

The masks for the targets in a chosen frame with the brightest target were first defined manually, then the background was automatically subtracted from the TPF in the kepextract task. Finally, the tasks of kepdraw and kepconvert were used to plot the extracted light curve and convert the data from FITS to ASCII. The original light curve plotted from the original ASCII data file was separated into several small parts to remove cosmic rays by filtering with a simple median filter. The converted and clarified ASCII data file has the three columns, i.e., Barycentric Julian Day (BJD), Simple Aperture Photometry (SAP) flux and the flux error. Note that SAP flux is expressed in electrons per second.

\footnotetext[1]{\scriptsize{IRAF (Image Reduction and Analysis Facility) is distributed by the National Optical Astronomy Observatory, which is operated by the Association of Universities for Research in Astronomy, under cooperative agreement with the National Science Foundation.}}

\subsection{Magnitude Conversion}

By convention, the transformation from the SAP flux f$_{kep}$ to the K2 magnitude Kp$_{2}$ can be approximated by the same formula used in the primary Kepler mission, i.e.,
\begin{equation}
\frac{f_{kep}}{1.74\times 10^{5}}\,=\,10^{-0.4(Kp_{2}-12)}
\end{equation}
During this transformation, we removed most of the single data points with obviously large scatter. In order to further verify this transformation, a simple comparison between the results from the PyKE and IRAF$^{1}$ was carried out. The task kepimages in PyKE was used to separate the TPF files into a series of time-series FITS images. Then, they were measured by using QPHOT in the aperture photometry package of IRAF. The final comparison indicate that there is only a small systemic difference between the two light curves. On the other hand, \citet{van14a} and \citet{van14b} have proposed a 'self-flat-fielding' (SFF) method to remove the effect of K2's trends. They defined a parameter of improvement percentage to describe their K2 light curves reduced by using the SFF method.  All the SFF corrected light curves are available to the community online at their own website. However, the SFF method does not always work well for the CVs. Table 1 lists the improvement percentages given by \citet{van14a} and \citet{van14b}, which are made by the SFF correction for all 15 CVs, of which 11 are less than 50\%, and 3 are negative. This means that the SFF method failed for three systems CSS130516:111236.7+002807 (hereafter CSS1112+0028), SDSS J113826.72+061919.5 (hereafter J1138+0619) and RZ Leo. Since QZ Vir, TW Vir and RZ Leo were observed in short cadence  mode which provides more information than their simultaneous LC data, we only discuss the SC light curves for these three CVs.

\subsection{Subtraction of Pseudo Periodic Signals}

There were two significant periodicities in all K2 observations. The first is a 6-hr periodic correction for the drift in the satellite pointing due to the loss of two reaction wheels. The second is a period at 48\,hrs due to thruster firings to remove momentum. This periodicity is only seen clearly in the three original light curves in short cadence plotted from the direct output data files of the PyKE, which show a nearly strict period of 3.9256\,day \citep{dai16}. Note that this period is the second harmonic of the period 48\,hr due to the plausible two models of thruster firing event. Both effects are common in the K2 field \citep[e.g.][]{dai16,hak15,kra15}. The pseudo periodic signals resulting from the jitter of the spacecraft in all of the extracted K2 light curves must be removed or minimized during the K2 data reduction. After several experiments, we found that the best mask covering 9 pixels can almost mitigate the impact of the 6-hr jitter of the target positions on the CCD, since the peak at the period of $\sim$\ 6\,hr completely disappeared in the periodograms of our deduced K2 data for all CVs discussed in this paper. Considering that the 48-hr pseudo periodic signal is a series of distinctly separated dips, it was easy to manually remove them from the three SC light curves in the following analysis.

\section{Results and Discussion}

The EPIC identifications for the targets in the K2 mission are listed in Table 1 along with the discovery name or standard variable star designation if available. Of the 15 CVs observed in K2-C0 and K2-C1, seven targets marked in Table 1 only show relatively featureless light curves in LC since they were very faint during the K2 observations. Hence, they will not be discussed. The basic parameters of the remaining eight CVs are listed in Table 2. Table 3 lists four transient events that were detected in the quiescent CVs with different subtypes. Since these events all have different profiles, durations, amplitudes and occurred at different times in the light curves of each campaign, they are likely not due to the data reduction process. While their origin is not clear as they are singular events in each case, it is likely that they are related to slight changes in accretion rates. Figure 1 shows the 3 CVs with normal outbursts observed in K2-C0. Figures 2 and 3 display the 5 CVs in quiescence and 2 CVs with superoutbursts.

\subsection{CVs in quiescence}

\subsubsection{EPIC\,202061317\,=\,1RXS\,J0632+2536\,=\,SDSS\,J063213.1+253623}

This CV was named 1RXS\,J0632+2536 (hereafter J0632+2536) in SIMBAD. \citet{rit03} and \citet{wat15} have classified this target as a DN with an unfiltered magnitude range of 12.4\,-\,15.3 in the updated catalogue of cataclysmic binaries and the AAVSO International Variable Star Index (VSX), respectively, based on the reports from the vsnet-alert archives \citep[e.g.][]{dem12a,dem12b}. It is also listed in several other catalogues in the infrared, optical, UV and X-ray bands, such as the 2MASS all-sky catalogue of point sources \citep{cut03}, WISE All-Sky Data Release \citep{cut12,cut14}, the USNO-A2.0 and B1.0 catalogues \citep{mon98,mon03}, the SDSS DR7 \citep{ade09}, the GALEX-DR5 (GR5) sources from AIS and MIS \citep{bia12}, and the ROSAT All-Sky Survey Faint Source Catalog \citep{vog00}. Despite these detections, it is still a poorly studied CV. GALEX measurements of FUV=18.42$\pm$0.05 mags, NUV=17.90$\pm$0.03 mags and the ROSAT X-ray count rate of 0.02\,cts/s are typical for CVs. Several DN outbursts in March 2009 and January 2012 are reported in the literature \citep[e.g.][]{kor12,mas12,ohs12a} and are the basis for its classification as a dwarf nova. The two discovery images obtained by a small telescope (13.5\,cm) in Ka-Dar Observatory located at the Mountain station of Kazan State University show obvious position deviation from the other observations \citep{kor12}. Both discrepant observations may be caused by their underestimated errors. All K2 measurements are of the brighter source in the DSS finding chart.

Considering that the SFF corrected light curve has only a negligible improvement of 3.6\% for this target, we only analyzed our reduced K2 light curve which is shown in the top panel of Figure 1. Besides a long-term periodic variation clearly seen in the zoomed light curve, there is a sharp decrease of system light occurring around BJD\,2456790 with an amplitude of $\sim$\ 0.25\,mag. After the normalization by removing the systematic variations from the original K2 light curve, a periodogram analysis clearly indicates two significant peaks at the period of 0.1621\,day (3.89\,hr) and 0.3240\,day (7.78\,hr). The method of phase dispersion minimization (PDM; \citep{ste78}), confirms the result of the periodogram. The top plot in Figure 4 shows that the period of 3.89\,hr has more power than that of 7.78\,hr. However, the phased light curve of the 35-day K2 data folded on the 7.78\,hr period, shown in the top panel of Figure 5, proves that this period is the correct one, showing two distinct peaks and dips during the orbit. A photometric period of $\sim$\ 0.32521\,day (7.81\,hr) was observed during a previous photometric campaign launched by the VSNET during an outburst that occurred in 2012 late January to early February \citep[e.g.][]{dem12a,dem12b,dub12a,dub12b}. The period of 3.89\,hr was also detected in the vsnet-alert report proposed by \citet{kat12} while other VSNET reports during the decline from outburst claim that it was variable \citep{ohs12b,ohs12c}. The 35-day K2 data clearly demonstrate the stability of the period of 7.78\,hr. It is very likely that this is the orbital period of the system and the longer period of 7.81\,hrs was a superhump period.

Several spectra were obtained over a timespan of one hour on 2015 February using the MODS spectrograph \citep{pog10} on the Large Binocular Telescope (LBT). The spectra cover 3400\,-\,7400\,\AA\ using 2 gratings with a resolution of about 2000. The summed spectra are shown in Figure 6 while the red spectra at the beginning and end of the timespan are shown in Figure 7. These spectra show many absorption features yielding an approximate spectral type of K5V, which is consistent with a long orbital period system. The strong doubled nature of the Balmer emission lines indicates a peak-to-peak separation of $\sim$\ 620\,km/s. During the hour timespan, the doubled lines changed to a single feature, as is typically viewed in DN as a disk structure such as a hot spot moves through the emission lines.

\citet{ohs12a} first regarded that the dips in the light curve are caused by eclipse, but the wide widths (about half an orbital cycle) and shallow depths (0.3\,mag and 0.1\,mag) are not consistent with eclipses. The light curve is similar to that of the long period (7.3\,hr) system TT Crt \citep{szk92}, where the modulations can be explained by ellipsoidal variations due to the K secondary if the mass ratio M$_{wd}$/M$_{sec}$ is $>$\,1 and the inclination is close to eclipsing ($\sim$\ 50\,-\,65\,degree). The stream flow over the disk for the high mass transfer rates expected for long period systems can create hot spots that contribute to the large amplitude of the ellipsoidal variation. However, it is surprising that the maximum brightness comes after the deepest minimum, whereas a hot spot is normally expected prior to the deep minimum. Spectroscopy throughout the orbit can provide confirmation of the correct scenario.

 \subsubsection{QZ Vir}
 
This is a well-known SU UMa type DN (former name of T Leo) with the explicit detection of several superoutbursts and their corresponding superhumps \citep[e.g.][]{lem93,kun94,kat97,pat05,kat09}. \citet{sha84} and \citet{kat09} obtained its orbital and superhump periods of 0.05882\,day (1.41\,hr) and 0.0604\,day (1.45\,hr), respectively. There are also other proposed identifications as a WZ Sge type star \citep{how99} or Intermediate Polar (IP). The latter identification was suggested by \citet{sha84} and \citet{vri04} because of a 414\,sec period observed in the XMM-Newton X-ray light curve that could be the spin period of the white dwarf. \citet{wen83} derived a mean cycle length of 420\,day for its superoutbursts based on four early eruptions reported in the literature. There are five known superoutbursts in 1993, 2005, 2007, 2008 and 2009 observed in the optical band \citep{kat09}, and one superoutburst in 1988 observed in the UV band \citep{ham04}. Compared with other typical WZ Sge type stars, the supercycle of QZ Vir is shorter, and more coherent.  

During the 80-day K2-C1 observations in short cadence, there was no outburst event, but a secular decline with an amplitude of $\sim$\ 1\,mag. AAVSO data show that a superoutburst began May 13 and lasted until about June 18 (about 16 days after the start of the K2-C1 observations). So, the decline is consistent with the end of the superoutburst. Quadratic and linear functions were used to normalize the original K2 light curve before and after the data gap of K2-C1, respectively. The normalized light curve is shown in Figure 2. Considering the long-cadence SFF corrected light curve had an improvement percentage of less than 10\%, the following discussions are based on our deduced light curve in short cadence, which shows quasi-periodic modulations with a mean amplitude of $\sim$\ 0.6\,mag. Note that this amplitude can occasionally reach up to over 1\,mag. Moreover, a special flare-like transient with a duration of $\sim$\ 8\,day and an amplitude of $\sim$\ 0.8\,mag marked by the dash rectangles in Figure 2 occurs around BJD\,2456878. The zoomed light curve of this transient shown in the top panel of Figure 8 further reveals that this event is also surrounded by other low-level modulations. Although the periodicity of this modulation cannot be visibly seen as in J0632+2536, the DFT result plotted in the second panel of Figure 4 using the program Period04 definitely indicates two significant peaks at the close periods of 0.0588\,day (1.41\,hr) and 0.0597\,day (1.43\,hr), which are nearly consistent with the previously derived orbital and superhump periods, respectively. The other peak at the period of 0.0294\,day (0.71\,hr) shown in this periodogram is the harmonic of the orbital period. In addition, we found that a group of peaks at the central period of 7.92\,hr may be related to the pseudo 6-hr period.

Since the decline at the beginning of the K2-C1 light curve is the end of a superoutburst, the fact that the detected period of 1.43\,hr derived from the whole K2 dataset is a little less than that obtained in the literature can be reasonably explained by the transition from the long superhump period to the short orbital period after the superoutburst. Since the data gap basically separates this K2 light curve into the decline and quiescent parts, the two periodograms for the two parts of light curves shown in the top two panels of Figure 9 further verify that the decline is the end of the superoutburst. The top panel indicates that the decay part of the superoutburst only shows the superhump period. The middle panel illustrates that the highest peak has been shifted from the superhump to the orbital period in quiescence. Besides these two periods, the quiescent light curve also includes many unknown periods (sum of these periods may be beat periods), which result in the complicated and irregular variations. Thus, we cannot obtain the distinct orbital and superhump modulations. Note that the peak at the superhump period does not disappear or become very weak in this plot. This may imply that the superhump period of 1.43 hr in QZ Vir is permanent and coexists with the orbital period. Although many periods were detected in quiescence, we cannot detect the spin period of the white dwarf at 414\,sec and its harmonics, as found by \citet{vri04}. The shortest period with a considerably weak power is 18.6\,min. Additionally, the periodogram for the flare-like transient shown in the (c) panel of Figure 9 reveals a curious feature in that both orbital and superhump periods nearly vanished, but the higher harmonic of the orbital period is significant. Furthermore, there is another notable peak at a new period of 2.67\,hr, which is not displayed in the other two panels.

\subsubsection{RZ Leo}

\citet{wol19} first classified RZ Leo as a possible nova in 1918 with a maximum at 10.5\,mag. But, the second outburst in 1984 and the strong double-peaked emission lines suggest that RZ Leo should be a member of WZ Sge-type stars \citep{mat85,cri85}. Then, \citet{gre85}, \citet{how88} and \citet{ish01} further proposed that RZ Leo should be a SU UMa star due to the detection of superoutbursts lasting 14 days and corresponding superhumps with a period of 0.07616(21)\,-\,0.0806(2)\,day and an amplitude of $\sim$\ 0.15\,-\,0.2\,mag. The secondary of RZ Leo is regarded as a normal red dwarf instead of a brown dwarf as implied by the typical WZ Sge stars \citep{men99,ish01}. In the literature, there were 7 recorded outbursts in 1918, 1935, 1952, 1976, 1984, 1989 and 2000 \citep[see][and references therein]{ish01}.  Hence, \citet{stu00} estimated that the supercycle of RZ Leo is approximately ten years. In quiescence, the photometric period of 0.0756(12)\,day derived by \citet{men99} is a little smaller than the spectroscopic period of 0.07651(26) day \citep{men01}. \citet{pat03} obtained similar photometric and spectroscopic periods with higher precision, which are 0.0760383(4)\,day and 0.0761(2)\,day, respectively. Both periods are smaller than the derived superhump period of 0.07868(19)\,day \citep{kat09,pat03}. 

In K2-C1, RZ Leo is another one of three CVs observed in short cadence. The SFF method failed to improve its K2 light curve due to the negative improvement percentage listed in Table 1. The nearly flat K2 light curve in short cadence shown in the middle panel of Figure 2 never shows any long-term variation, but does have regular and low-amplitude oscillations ($\sim$\ 0.5\,mag on average) in quiescence. The Lomb-Scargle periodogram plotted in the middle panel of Figure 4 indicates two notable periods of 0.03801\,day (0.91\,hr) and 0.07603\,day (1.82\,hr). This periodogram is similar to Figure 8 of \citet{pat03}. Additionally, the PDM method also confirmed this result. Hence, like J0632+2536, the period of 0.07603\,day with less power should be an authentic period, which agrees with the previous result derived from the double-hump photometric light curves \citep{pat03}. Furthermore, based on the 82-day K2-C1 data in short cadence, this period can be improved by the O-C technique due to the high time resolution of the K2 data. Considering that the irregular profile of the second hump seriously distorts the two dips, the maximum time of primary hump should be the best periodic signal for our O-C analysis. The first maximum time and the period obtained from the periodogram were set to be an initial epoch and period, respectively. A primary ephemeris based on a time baseline over 1000 cycles can be used to rephase the short-cadence K2 data.  However, when binning this phased light curve with a phase resolution 0.001, there is a small zero-drift in phase $\sim$\ 0.015 due to the incorrect epoch. The new ephemeris after removing this zero-drift effect was deduced as:
\begin{equation}
T_{max}\,=\,BJD\,2456808.23662(2)\,+\,0.07602997(4)\,E,
\end{equation}
with a variance of $7.4\times10^{\rm -7}$\,day. The precision of the period shown in Equation (2) is about one order of magnitude higher than the period derived by \citet{pat03}. Figure 10 displays all the O-Cs of the maximum times of primary hump based on the ephemeris Equation (2). By using the new derived period listed in Equation (2) to fold the linearly normalized K2 light curve, we found that the phased light curve shows very large scatter, $\sim$\ 0.43\,mag on average. Compared with the phased light curve derived by \citet{pat03}, our smooth binned light curve with higher phase resolution shown in the middle panel of Figure 5 distinctly illustrates more details of the orbital light curve including a round bottom at the primary dip and the round tops at the two peaks. A flat bottom at the second dip lasting phase 0.05 ($\sim$\ 5.5\,min) is a consistent feature.

This phased light curve illustrates a brightening transient event with an amplitude of $\sim$\ 0.6\,mag, which is also displayed in the middle panel of Figure 8. Both plots clearly demonstrate that this event takes place at BJD\,2456845.1, which is coincident with the original second hump. In the phased light curve, the top part of this transient event between phase 0.3 and 0.9 obviously changes the normal second hump. The complete profile of this transient shown in the plot of time series indicates that its duration is $\sim$\ 1.88\,hr, which is a little longer than the orbital period, and it is composed of a rapid rise with a rate of $\sim$\ 1.6\,mag\,hr$^{-1}$, a flat peak with a rate of $\sim$\ 0.05\,mag\,hr$^{-1}$ and a slow decline with a rate of $\sim$\ 0.5\,mag\,hr$^{-1}$. This event looks like a mini DN outburst with a faster and lower amplitude. In addition, there is a weak rebrightening event occurring in the decay with a typical luminosity level compared with the other primary peaks. This event is consistent with the ephemeris of Equation (2). So it is a normal primary hump and is marked by a short vertical line in the middle panel of Figure 8.

\subsubsection{EPIC\,201525706\,=\,WD\,1144+011\,=\,SDSS\,J114633.93+005510.3}

LBQS\,1144+0111, an alias of this target, was first observed in the Large, Bright QSO Survey (LBQS, \citet{fol87,fol89}), and classified as a DA+dMe binary, which is composed of a DA type white dwarf and a normal M dwarf star with strong emission features \citep{ber92}. Additionally, it was listed in a catalog of spectroscopically identified white dwarfs in the first data release of the SDSS (g=18.09\,mag, \citet{kle04}). Although \citet{tap04} did not detect any variability in later time-series photometry, it was claimed as a cataclysmic variable named WD\,1144+011 in SIMBAD. However, its optical spectrum obtained by \citet{ber92}, shows many absorption lines and a low blue continuum, which is atypical for a normal CV. Thus, its CV identification is still doubtful. 

The continuous K2 light curve in long cadence covering 80 days shows an orbital modulation but no outburst or transient event. There is a long-term rise of the system light after the data gap of K2-C1. Therefore, we used linear and quadratic functions to normalize the K2 light curves before and after the data gap, respectively. Note that a small transient event marked by the dash rectangle in the bottom panel of Figure 2 appears at the end of the K2 light curve around BJD\,2456888. The zoomed light curve of this transient shown in the bottom panel of Figure 8 illustrates that it is a short-timescale event lasting $\sim$\ 0.5\,day, which involves a temporary disappearance of the orbital modulation. Since there is no notice in the data release notes for K2-C1 nor any similar variation in the other CVs in K2-C1, this appears to be a real feature of WD\,1144+011. The orbital modulation with an average amplitude of 0.03\,mag reappears after this transient event. By using the Lomb-Scargle periodogram and PDM methods, a highly significant period of 0.40859\,day (9.806\,hr) and its harmonics are found based on the normalized data. Since the K2 light curve cannot be used for a reliable O-C analysis like RZ Leo due to the sparse data points, we attempted to use a sinusoidal function to fit the normalized light curve according to the Levenberg-Marquardt algorithm. The best least-square fitting curve shown in Figure 11 can be deduced as follow,
\begin{equation}
M_{nor}=-9.9650(2)\times10^{-4}+1.905555(2)\times10^{-2}\,SIN[2\pi\frac{T_{BJD}}{0.408680655(1)}\,-\,2.524317(2)],
\end{equation}
where M$_{nor}$ and T$_{BJD}$ are the normalized magnitude and the modified BJD (i.e., T$_{BJD}$\,=\,BJD\,-\,2456000), respectively. The best fitting period of 0.408680655(1)\,day in Equation (3) basically accords with the primary period shown in the periodogram. Since the SFF improvement percentage is 10.6\%, we attempted to phase the SFF corrected light curve based on the period in Equation (3) and obtained a nearly equal phase gap at a frequency of 20\,cycle/day (i.e., $\sim$\ 19\,min interval) shown in the (a) row of Figure 12. Although our normalized light curve folded by the same period shown in the (c) row presents a similar gap, the zoomed-in light curve in the right-side panel of the (c) row indicates that each section of the phased light curve is smoother than that of the (a) row. In addition, by using the DFT method (Period04 program), we obtained a new period of 0.408741\,day (9.81\,hr), which only has a small difference 5.2\,sec from the period listed in Equation (3). In spite of this small discrepancy, the new phased SFF corrected light curve shown in the (b) row illustrates a distinctly different pattern of a sine-like modulation with a phased period of 0.02\,day (i.e., $\sim$\ 29.4\,min) superimposed on the original phased light curve without any gap. This modulation is totally invisible in the new phased light curve for the normalized data, which only displays smooth scatter. Note that the modulation period of 29.4\,min presented in the (b) row is exactly the same as the sampling rate of the K2 in LC and one 20th of the folding period of 9.81\,hr. Therefore, this sine-like modulation derived from the SFF corrected data is likely not real since it could be removed after a simple normalization. The 19-min phase gaps shown in the a) and c) rows may be caused by the inaccurate folding period. The binned phased light curve shown in the bottom panel of Figure 5 has a slow rise and fast decline. The long period and the light curve shape imply this system is likely a pre-CV with the modulation caused by the changing view of the M star that is heated by the hot white dwarf. Time resolved spectroscopy throughout the 9.81\,hr period should show noticeable changes in the emission lines.

\subsection{CVs in outburst}

\subsubsection{EPIC\,202061315\,=\,USNO-B1.01144-00115322}

This object named USNO-B1.01144-00115322 (hereafter USNO\,01144) was first listed in the USNO-B1.0 catalogue \citep{mon03}. \citet{kry09} claimed that it showed a DN outburst with a duration of\ $>$\ 7\,day and an amplitude of\ $>$\ 4\,mag as observed at KSU Astrotel Observatory in November, 2009. Its optical spectra on 2009 November 19 obtained by \citet{pej09} show broad Balmer absorption lines with a narrow emission core and a weak H$\alpha$ emission line in the blue and red wavelengths, respectively. In addition, the continuum in the blue end is obviously stronger than that in the red end. 

During the observations with K2-C0 in long cadence, a brightening event was detected around BJD\,2456770. Although the K2 light curve shown in Figure 1 was incomplete due to missing the beginning part of the outburst, it was enough to confirm that this is a typical normal outburst with an outburst amplitude of $\sim$\ 3.5\,mag, which further supports the previous classification of DN. However, the SFF corrected light curve with the significant improvement percentage 209.6\% never recorded this brightening event since the SFF data just began from BJD\,2456772.109, close to the end of this normal outburst. Based on the systematic difference between our normalized SAP flux and the SFF corrected data, the SFF corrected light curve was replotted in the middle panel of Figure 1 with the same scale as our deduced K2 light curve. We found that the large scatter with an amplitude of $\sim$\ 1\,mag has been almost eliminated by the SFF method. Especially, the sudden introduced large scatters at the end of the smooth decay of normal outburtst can be naturally explained by its faint quiescence. The plot of the differences between the two light curves illustrate an obvious deviation lasting $\sim$\ 5\,days at the end of the decay section, which is caused by the very flat SFF light curve. Considering that the diagram of the differences between the two light curves in quiescence is basically flat as shown in the middle panel of Figure 1, the normal outburst with a smooth and consecutive profile presented in our deduced K2 light curve is authentic. Thus, a new and complete K2 light curve of this object can be composed of our deduced part in outburst and the SFF corrected part in quiescence. A period search based on this complete K2 light curve only gives the null result. 

\subsubsection{UV Gem}

The three superoutbursts recorded in 2003, 2008 and 2011 accompanied with superhumps definitely classify UV Gem as a SU UMa-type star with a long superhump period \citep[e.g.][]{kat01,kat09}. The O-C diagram of the superhumps derived by \citet{kat13} show a downward quadratic curve  covering these three superoutbursts with an extremely negative decline rate of -5.34$\times$10$^{-4}$ on the basis of the superhump period of 0.0936\,day derived from the superoutburst in 2003. \citet{kat01} estimated that the cycle length of the normal outbursts is $\sim$\ 13.5\,day, but there is not enough data to derive the supercycle length \citep{otu13}. Until now, only one quiescent spectrum with a weak H$\alpha$ emission line and typical continuum covering 4600\,-\,9000 \AA\ was obtained by \citet{zwi94}. Thus, its orbital period is still unclear \citep{rit03} 

Fortunately, a complete normal outburst lasting 5 days with an amplitude of around 3\,mag occurring around BJD\,2456788 was detected in K2-C0. Considering that the SFF method successfully improved the quiescent light curve of USNO\,01144, the SFF corrected light curve was checked but there was a very low improvement percentage 0.8\% for UV Gem. In order to clearly demonstrate the coherence of the two light curves, they are replotted into the same frame and their almost flat differences are shown in the bottom panel of Figure 1 except for a systematic shifting $\sim$\ 0.1\,mag from BJD\,2456783 to BJD\,2456794, which totally covers the whole outburst. Like USNO\,01144, the flat differences imply that the normal outburst detected in our deduced light curve is also real. In addition, the large scatter and the quasi-periodic jitters at the peak and decay parts of the outburst light curve clearly appeared in the SFF corrected light curve. The two Lomb-Scargle periodograms for our deduced data and the SFF corrected data shown in the UV Gem (a) and (b) panels of Figure 13 demonstrate that the known 6-hr pseudo periodic signal caused by the regular thrusting event of the Kepler satellite is not eliminated in the SFF data, but is removed from our deduced data. Additionally, the two significant peaks at the period of 0.086089\,day ( 2.07\,hr) and 0.088342\,day (2.12\,hr), respectively, appeared in the UV Gem (a) panel of Figure 13. Both periods are less than the orbital period of 0.0895\,day (2.15\,hr) listed in the updated CVs catalog (RKcat\,7.21, \citet{rit03}), that was estimated from the superhump period of 0.0935\,day (2.24\,hr) using an empirical relation given by \citet{sto84}. Hence, the K2 data allow the first determination of a plausible orbital period for UV Gem. Moreover, there is still a weak but analogous frequency feature displayed in the UV Gem (b) panel of Figure 13. Considering that this periodogram is similar to QZ Vir, the longer 2.12\,hr period with less power may be the remaining superhump period after an unrecorded superoutburst. However, the phased light curve cannot be obtained due to the sparse K2 data available in long cadence.

\subsubsection{EPIC\,210282487\,=\,1RXS\,J111236.7+002807\,=\,CSS1112+0028\,=\,SDSS\,111236.7+002807}

In SIMBAD, this target named 1RXS\,J111236.7+002807 is classified as an X-ray source as it first appeared in the ROSAT All-Sky Survey Faint Source Catalogue \citep{vog00}, and has a typical CV count rate of 0.02368. However, it was regarded as a galaxy (g\,=\,22.46\,mag) in the SDSS DR 9 in spite of a position deviation of $\sim$\ 14.8\,arcsec compared with the coordinates in SIMBAD. Recently, the CRTS classified this target as a CV due to the detection of several transient variations with a nearly 4\,mag amplitude (i.e., 17\,-\,21\,mag) in low time resolution data obtained during the last ten years \citep{dra09}. Hence, the faint g-band magnitude listed in the SDSS catalog corresponds to quiescence during the SDSS observations.

A typical superoutburst lasting 17 days with a weak precursor and an amplitude of $\sim$\ 5.6\,mag was detected in the long cadence K2-C1 data. The converted K2 magnitude is consistent with the previous observations of the CRTS and the SDSS. Although the scatter in the quiescent light curve is not apparent in the upper panel of Figure 3, it is obviously present in the magnitude plot (i.e., the log graph) since quiescence may reach the detection limit of the satellite. Considering that the SFF method only provides a negative improvement percentage -19.5\%, we cannot expect that the scatter in quiescence can be reduced by the SFF method. In order to visually display the superhumps on the plateau of the superoutburst, we used the dotted lines to connect the adjoining data points due to the low time resolution in long cadence. The zoomed plots in SAP flux and Kp$_{2}$ are shown in the (a) and (b) panels of Figure 14, respectively. Both diagrams further demonstrate that it is hard to reliably detect the superhump from such sparse time series, which only has 4\,-\,5 data points in each superhump cycle at most. In spite of this, we attempted to use a linear function to normalize the plateau in magnitude, which can be described as follow,
\begin{equation}
Kp_{2}\,=\,-59.813(\pm0.499)\,+\,0.088639(\pm0.000574)\,T_{BJD}.
\end{equation}
This can be converted into an exponential function for describing the plateau in SAP flux by using Equation (1). However, the normalized plateau in magnitude shown in the (c) panel of Figure 14 shows a quasi-sinusoidal variation with an amplitude of $\sim$\ 0.05\,mag and a period of $\sim$\ 7.6\,day except for the small deviation at the end of the plateau. After a further normalization by subtracting this sine curve, a flat plateau shown in the (d) panel of Figure 14 was obtained. Based on this final normalized data, a Lomb-Scargle periodogram shown in the bottom panel of Figure 4 reveals two significant peaks with comparable power at 0.060137\,day (1.44\,hr) and 0.30068\,day (0.72\,hr). However, the PDM method cannot find any period from the same data. Considering that the minimum orbital period for hydrogen-rich CVs is $\sim$\ 70\,min \citep{war03}, it is reasonable to conclude that the smaller period of 0.30068\,day is only a harmonic, but the superhump period of this object is the other period of 0.060137\,day due to the lack of peaks at longer periods. This implies that this object is a short-period SU UMa type star. Although the superhump period derived from this sparse K2 data may be not very accurate, the K2 light curve in long cadence can definitely confirm the detection of a typical superoutburst and the corresponding superhump with a plausible period of 1.44 hr. This means that CSS1112+0028 is a SU UMa type DN close to the period minimum. Note that the known 6-hr pseudo period is also not visible in the periodogram. Further followup observations in higher time resolution are necessary to precisely certify its superhump and orbital periods.

\subsubsection{TW Vir}

This is a well-known UG Gem type DN since 1932 \citep{oco32}. \citet{sha83} first obtained its orbital period of 4.38\,hr, the orbital inclination 43$^{\circ}$ and the mass ratio of 0.44. \citet{ak02} claimed that there were 7872 observations including 62 normal outbursts for TW Vir from 1955 to 1997. Based on these observations, they derived that the average outburst duration and the quiescent interval of TW Vir should be 6 and 22 day by setting the two reference lines at 12.2 and 13.5\,mag. Besides this optical photometry, there were some detailed UV and IR observations. It's UV spectrum in outburst taken by the IUE shows a P Cygnic profile of the CIV\,1550 line, which indicates a strong accretion disk wind \citep{cor82,ham07}. \citet{szk85} analyzed other IUE spectra when TW Vir switched to quiescence, and found that the CIV line returned back to a normal emission line with a larger equivalent width (145 \AA) than the Balmer lines. Additionally, \citet{mat85} obtained a sinusoidal IR light curve with a period of 2.2\,hr (i.e., one-half the orbital period) and a half amplitude of 0.063\,mag. They estimated that the secondary of TW Vir is a M3V red dwarf of 0.4 M$_{\odot}$.

Like QZ Vir and RZ Leo, TW Vir was observed in short cadence with K2-C1. The K2 light curve shown in the bottom panel of Figure 3 recorded two incomplete normal outbursts and one completed superoutburst lasting 16 days with a typical precursor on the rise to superoutburst. The amplitude of this superoutburst is $\sim$\ 4.5\,mag, which is 0.5\,mag larger than the following two normal outbursts. The quiescent interval between them is less than 26 days, which is consistent with the average result derived by \citet{ak02}. The amplitude and the duration of this superoutburst of TW Vir are 1.1\,mag and 1\,day shorter than those of CSS1112+0028, respectively. Considering that the improvement percentage by the SFF method is 16.9\% for the light curve in long cadence, the SFF corrected data and our reduced data were replotted in the same diagram like USNO\,01144 and UV Gem. Inspection of the bottom panel of Figure 3 indicates that the obvious deviations between the two light curves are at the beginning of the time series including the plateau of the superoutburst, where the SFF data present larger scatter. But, the remaining parts of the two light curves are similar. The two panels illustrate that the SFF light curve undergoes an unexpected break-off point near the end of the plateau. A periodogram for the plateau of the superoutburst based on the SFF data is shown in the TW Vir (b) panel of Figure 13. A peak at the period of 6\,hr is the only significant feature, and there is not any signal around the orbital period. For comparison, the periodogram of our deduced data in the same data range shown in the TW Vir (a) panel indicates that the plateau also seriously suffers from the effect of the pseudo 6-hr period. But, there are some weak peaks around the orbital period of 4.38\,hr.

\section{Conclusion}

The extracted light curves of the 15 CVs observed in the K2-C0 and K2-C1 fields reveal interesting details. J0632+2536, QZ Vir, RZ Leo and SDSS\,J1146 were observed only in quiescence while USNO\,01144, UV Gem, CSS1112+0028 and TW Vir also show outburst and/or superourtbursts. The major results from these light curves are listed below.

J0632+2536: A consistent period of 7.78\,hr derived from the normalized data is likely the orbital period, and the 7.81\,hr period previously during outburst is likely the superhump period. The phased light curve based on this period shows a regular orbital modulation with two distinct humps and dips superimposed on a long-term variation of system light. Based on the large width and shallow depth of the two dips, combined with our optical spectra showing a prominent K5 star, we regard this object as a moderate inclination CV with ellipsoidal variations in combination with one or more hot spots. A decrease of system light with an amplitude of 0.25\,mag and a duration of 2.5\,day was detected during the latter half of the K2-C0 light curve, likely indicating a change in accretion rate.

QZ Vir: The short cadence K2-C1 data demonstrate the coexistence of orbital and superhump periods in quiescence, which are similar to those in the literature. A declining light curve with an amplitude of $\sim$\ 1\,mag and a duration of $\sim$\ 25\,day before the data gap of K2-C1 is connected to the decay of a superoutburst and only one significant peak at the superhump period appears in the periodogram. Additionally, a flare-like transient event in quiescence with an amplitude of $\sim$\ 0.8\,mag and a duration of $\sim$\ 8\,day was apparent and surrounded by low-level modulations. During this transient event, the orbital and superhump periods vanished, but the 0.71\,hr harmonic of the orbital period and a new period of 2.67\,hr were apparent. Although the spin period of the white dwarf at 414\,sec detected in X-ray and its harmonics were not detected in the K2 light curve, multi-period variations were present.

RZ Leo: A new orbital ephemeris based on the 82-day K2-C1 data further confirms the period derived by \citet{pat03}. However, the phased light curve shows a large average scatter of $\sim$\ 0.43\,mag. After binning into 0.001 bins and using our refined ephemeris, we obtained a smooth phased light curve, with primary and secondary minima. Additionally, an unusual brightening event with an amplitude of $\sim$\ 0.6\,mag and a duration of $\sim$\ 1.88\,hr is apparent in one orbit.

WD\,1144+011: The K2 light curve in long cadence reveals a strong 5\% modulation of the light curve at a period of 9.81\,hrs. The analysis is complicated by a beat between this period and the LC sampling rate. It appears that this system is likely a pre-CV, with the modulation resulting from the changing view of the M dwarf which is heated by the white dwarf. The light curve shows a peculiar transient event lasting about a half day when the regular modulation almost disappears.

USNO\,01144: An incomplete normal outburst with a typical amplitude of $\sim$\ 3.5\,mag was detected at the beginning of K2-C0. This event was totally missed by the start times of the SFF corrected data. However, the SFF method greatly reduces the large scatter with an amplitude of $\sim$\ 1\,mag during quiescence time. This outburst event detected by K2-C0 classifies this target as a DN.

UV Gem: A complete and typical normal outburst with a duration of 5\,day and an amplitude of 3\,mag was observed in K2-C0. The SFF corrected and our deduced light curves are basically consistent. The two close periods of 0.086089\,day and 0.088342\,day shown in the periodogram are likely the superhump and orbital periods.

CSS1112+0028: This is a faint CV with large scatter in quiescence. The K2-C1 light curve in long cadence illustrates a typical superoutburst with an amplitude of $\sim$\ 5.6\,mag and a duration of $\sim$\ 17\,day. After the linear normalization, the plateau of the superoutburst shows a sine-like variation with a period of $\sim$\ 7.6\,day and an amplitude of $\sim$\ 0.05\,mag. After removing this sine-like variation, the final flat normalized plateau clearly reveals a superhump period 1.44\,hr. This means that this target is a new SU UMa type star close to the period minimum.

TW Vir: A superoutburst with a duration of $\sim$\ 16\,day and an amplitude of $\sim$\ 4.5\,mag followed by two incomplete normal outbursts was detected by K2-C1 in short cadence. In addition, the K2-C1 data indicates that the quiescent interval is close to 26 days.

\acknowledgments

This work was partly supported by CAS "Light of West China" Program and 1-year Visiting Scholar Program of The Chinese Academy of Sciences during a visit to Professor Paula Szkody in the University of Washington. PS acknowledges support from NSF grant AST-1514737. MK and PG acknowledge support from the Naughton Foundation and the UCC Strategic Research Fund.

\begin{table}
\caption{All 15 CVs observed in K2-C0 and K2-C1.}
\begin{center}
\begin{tabular}{lccccccc}
\hline\hline
&CV name$^{a}$ & EPIC\,& R.A. & Dec. & Duration & Cadence$^{b}$ & SFF\\
&&& J2000 & J2000 & day &&\%\\
\hline
K2-C0&&&&&&&\\
\hline
&CI Gem$^{c}$ & 202061316 & 06:30:05.86 & +22:18:50.7 & 33.1 & LC & 33.2\\
&KZ Gem$^{c}$ & 202061320 & 06:53:02.76 & +16:39:50.3 & 33.1 & LC & 490.9\\
&MLS0656+2514$^{c}$ & 202061321 & 06:56:05 & +25:14:44 & 33.1 & LC & 99.3\\
&J0632+2536 & 202061317 & 06:32:13.09 & +25:36:23 & 36.3 & LC & 3.6\\
&USNO\,01144 & 202061315 & 06:26:57.69 & +24:29:07.2 & 37.1 & LC & 209.6\\
&UV Gem & 202061318 & 06:38:44.0 & +18:16:12 & 36.3 & LC & 0.8\\
\hline
K2-C1&&&&&&&\\
\hline
&J1125-0016$^{c,d}$ & 201445903 & 11:25:55.72 & -00:16:38.5 & 80.0 & LC & 325.8\\
&MLS1115+0510$^{c}$ & 210282488 & 11:15:37.1 & +05:10:01 & 79.3 & LC & 22\\
&MLS1137+0042$^{c}$ & 210282489 & 11:37:51.0 & +00:42:17 & 80.1 & LC & 1.2\\
&QZ Vir & 201683812 & 11:38:26.82 & +03:22:07.0 & 80.0 & SC \& LC & 6.4$^{e}$\\
&CSS1112+0028 & 210282487 & 11:12:36.7 & +00:28:07 & 82.1 & LC & -19.5\\
&RZ Leo & 201585290 & 11:37:22.27 & +01:48:58.5 & 82.2 & SC \& LC & -1$^{e}$\\
&J1138+0619$^{c}$ & 201850883 & 11:38:26.73 & +06:19:19.5 & 80.0 & LC & -0.7\\
&WD\,1144+011 & 201525706 & 11:46:33.94 & +00:55:10.4 & 80.0 & LC & 10.6\\
&TW Vir & 201185922 & 11:45:21.16 & -04:26:05.6 & 80.1 & SC \& LC & 16.9$^{e}$\\
\hline\hline
\end{tabular}
\end{center}
\footnotesize{Note, $^{a}$ the abbreviated names are used for the objects without standard variable star designations. $^{b}$ LC and SC refer to the long and short cadence observations, respectively. $^{c}$ the faint CV is not discussed in this paper. $^{d}$ J1125-0016 refers to the CV 2dFJ112555-001639. $^{e}$ the SFF correction is only for the LC data.}
\end{table}

\clearpage

\begin{table}
\caption{Basic Parameters of the 8 CVs.}
\begin{center}
\begin{tabular}{lcccccc}
\hline\hline
&Name & Campaign & P$_{orb}$ & Magnitude & Classification & Ref.\\
&&& hr & V / SDSS g &&\\
\hline
Quiescence&&&&&&\\
\hline
&J0632+2536 & C0 & 7.82 & 12.6 - 17.9 & DN & 1\\
&QZ Vir (T Leo) & C1 & 1.41 & 10 - 15.9 & SU UMa / IP$^{a}$ & 1\\
&RZ Leo& C1 & 1.83 & 12.1 - 19.3 & SU UMa / WZ Sge & 1\\
&WD\,1144+011 & C1 & -- & 18.09 & CV$^{b}$ / DA+dMe & 2, 3\\
\hline
Outburst&&&&&&\\
\hline
&USNO\,01144 & C0 & -- & 16 - 20 & DN & 4\\
&UV Gem & C0 & 2.15 & 14.7 - 18.8 & SU UMa & 1\\
&CSS1112+0028 & C1 & -- & 17.55 - 22.1 & CV$^{c}$ & 5\\
&TW Vir & C1 & 4.38 & 12.0 - 16.3 & U Gem & 1\\
\hline\hline
\end{tabular}
\end{center}
\footnotesize{Note, $^{a}$ the unconfirmed classification. $^{b}$ the CV classification is only shown in SIMBAD. $^{c}$ from the CVs catalogue of CRTS. References: (1) \citet{rit03}; (2) \citet{ber92}; (3) \citet{tap04}; (4) \citet{kry09}; (5) \citet{dra09}.}
\end{table}

\begin{table}
\caption{Transient events of the 4 CVs in quiescence.}
\begin{center}
\begin{tabular}{ccccc}
\hline\hline
Name & Campaign & T$_{event}^{a}$ & Duration & Amplitude \\
&& BJD & hr & mag\\
\hline
J0632+2536 & C0 & 2456790 & 60 & 0.25\\
QZ Vir (T Leo) & C1 & 2456878 & 192 & 0.8\\
RZ Leo& C1 & 2456845 & 1.88 & 0.6\\
WD\,1144+011 & C1 & 2456888 & 12 & --\\
\hline\hline
\end{tabular}
\end{center}
\footnotesize{Note, $^{a}$ the occurring time of transient event}
\end{table}

\clearpage

\begin{figure}
\centering
\includegraphics[width=14.0cm]{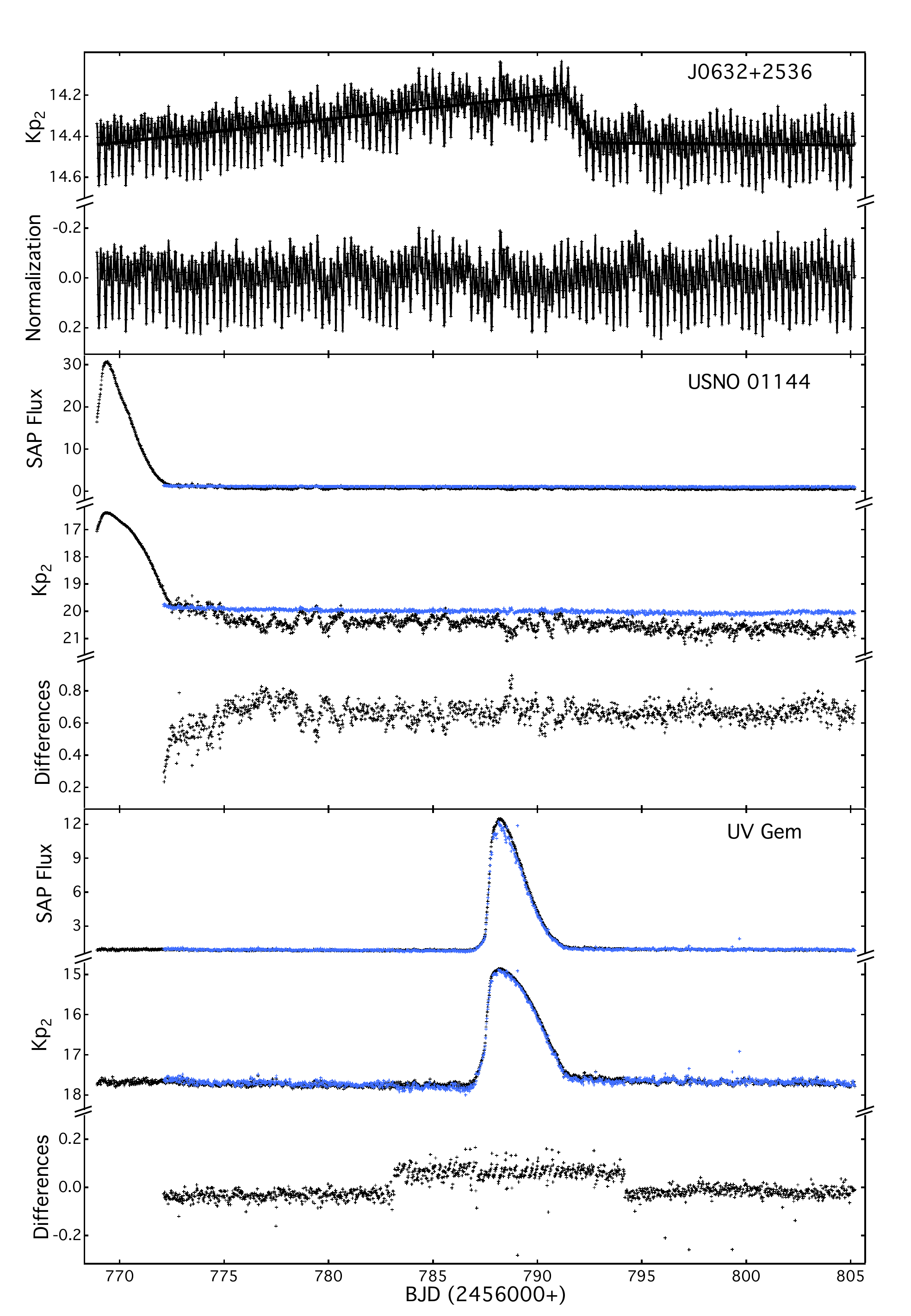}
\caption{\small{The three CVs observed in K2-C0. The top panel includes the normalized light curve of J0632+2536 without any long-term variation of system light. The blue light curves in the middle and bottom panels of USNO\,01144 and UV Gem refer to their rescaled SFF corrected light curves, respectively. And the difference between the blue and dark light curves in Kp$_{2}$ are plotted below. The SAP Flux of USNO\,01144 and UV Gem are in the units of 10$^{2}$ and 10$^{3}$ e$^{-}$s$^{-1}$, respectively.}} \label{Figure 1}
\end{figure}

\begin{figure}
\centering
\includegraphics[width=14.0cm]{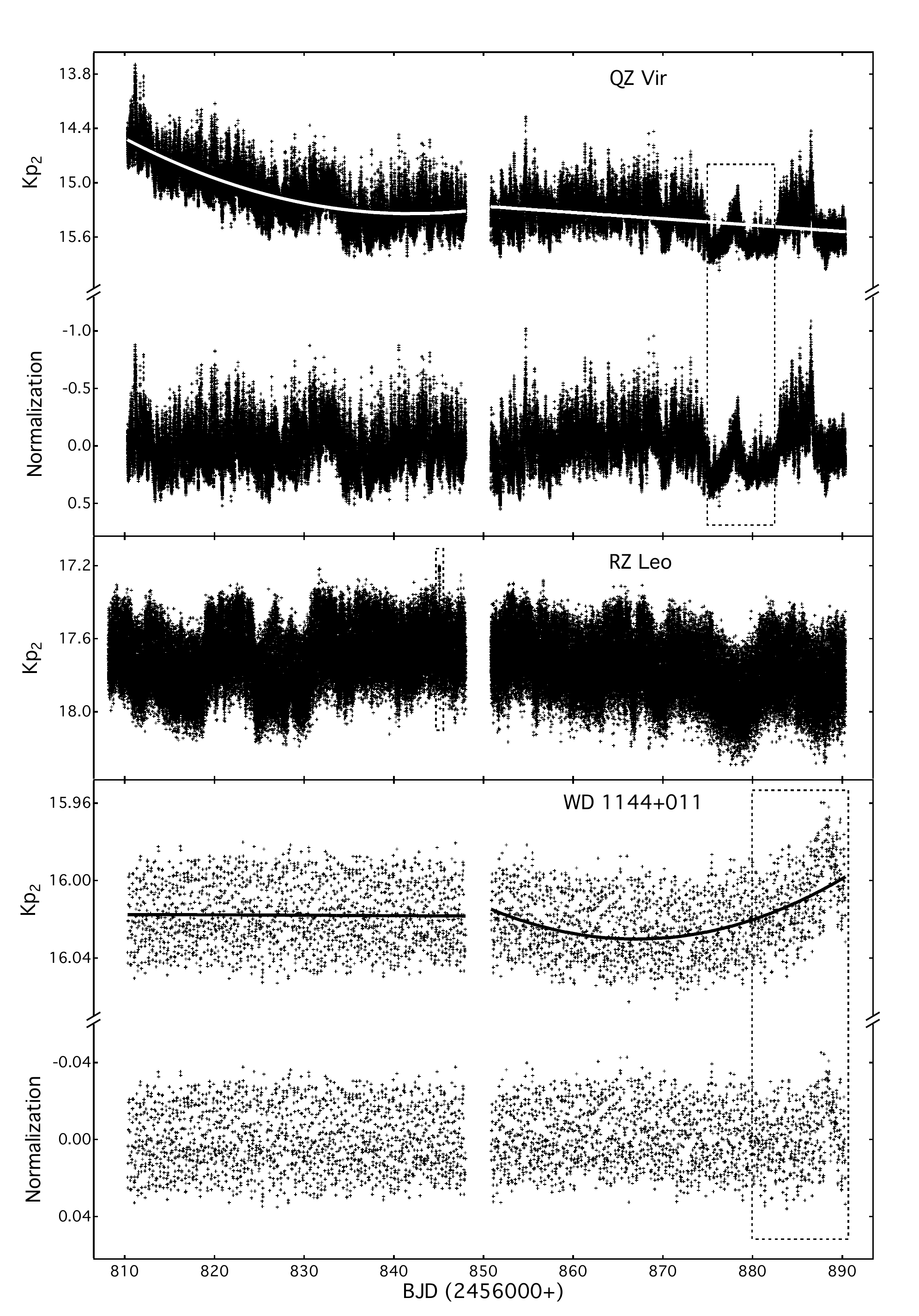}
\caption{\small{The quiescent light curves of the three CVs observed in K2-C1. The dash rectangles mark three special transient events, which are plotted in Figure 8. Two panels of QZ Vir and WD\,1144+011  include the normalized light curves.}} \label{Figure 2}
\end{figure}

\begin{figure}
\centering
\includegraphics[width=14.0cm]{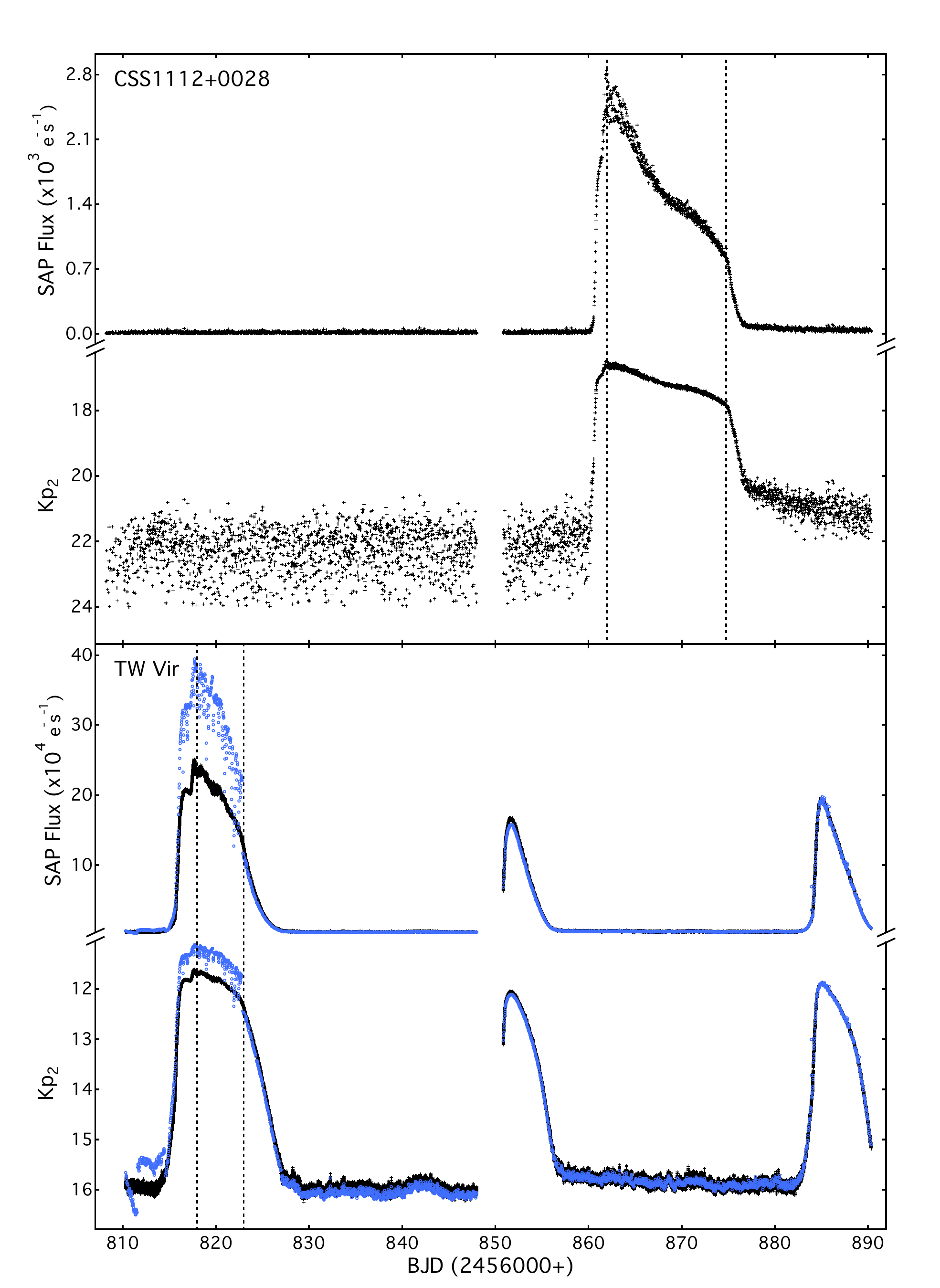}
\caption{\small{Two CVs with the outburst events observed in K2-C1. The light curves between two vertical dash lines in two panels are defined as the plateaus of superoutbursts of CSS1112+0028 and TWVir. The former plateau is zoomed in Figure 14. The latter one in Kp$_{2}$ was directly used to carry out a period search. Like Figure 1, the blue light curves in the bottom panel are the rescaled SFF corrected light curves of TW Vir.}} \label{Figure 3}
\end{figure}

\begin{figure}
\centering
\includegraphics[width=14.0cm]{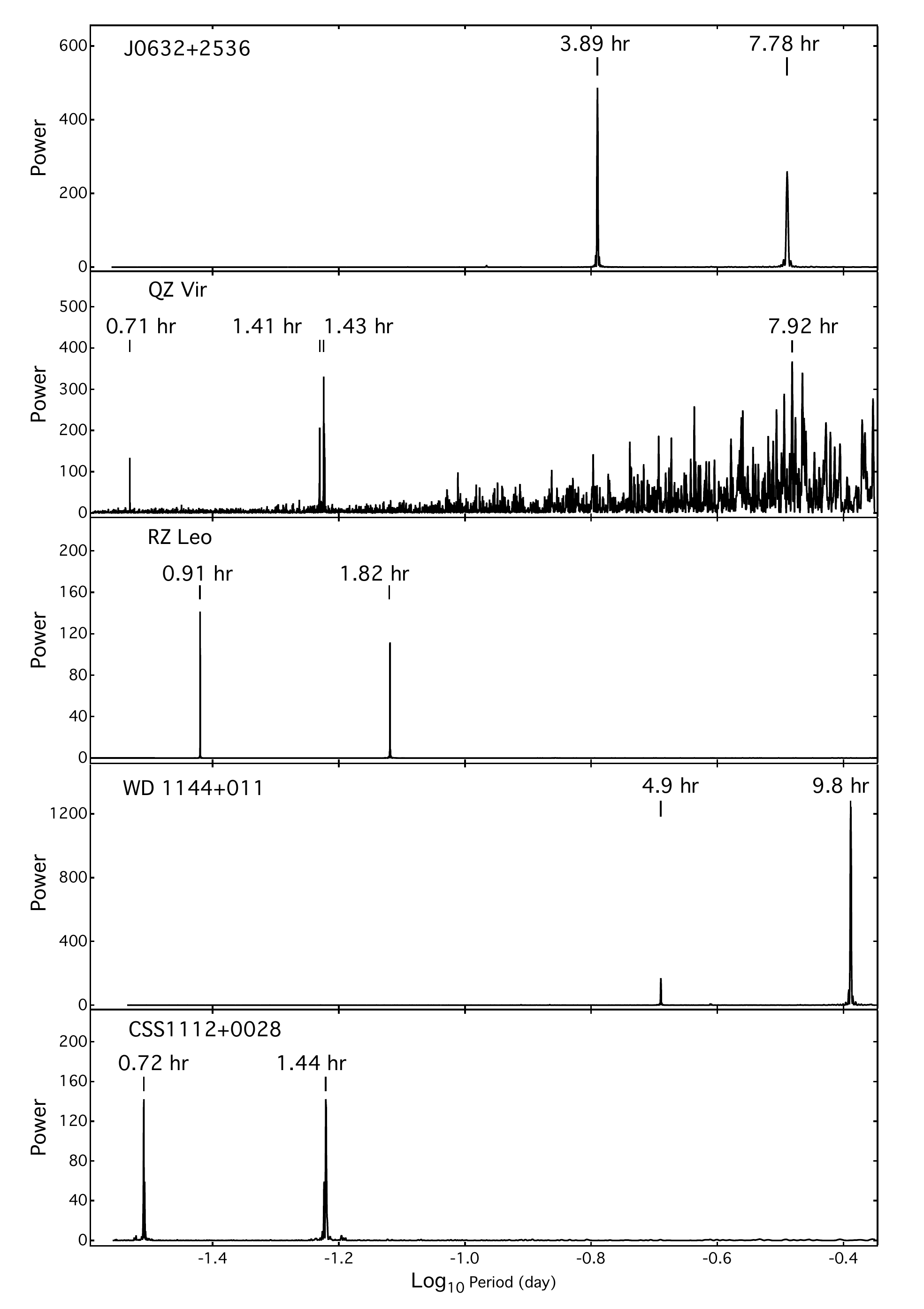}
\caption{\small{Period searches of the five normalized K2 light curves based on the Lomb-Scargle periodogram method. The significant periods and their harmonics are marked by the short vertical lines. The upper four panels present the periodograms for the complete and normalized K2 light curves in quiescence shown in Figure 1 and 2. The bottom panel illustrates the periodogram of CSS1112+0028 for the normalized plateau of superoutburst shown in the (d) panel of Figure 14. None of the diagrams show any notable peak at the known false 6-hr period.}} \label{Figure 4}
\end{figure}

\begin{figure}
\centering
\includegraphics[width=16.0cm]{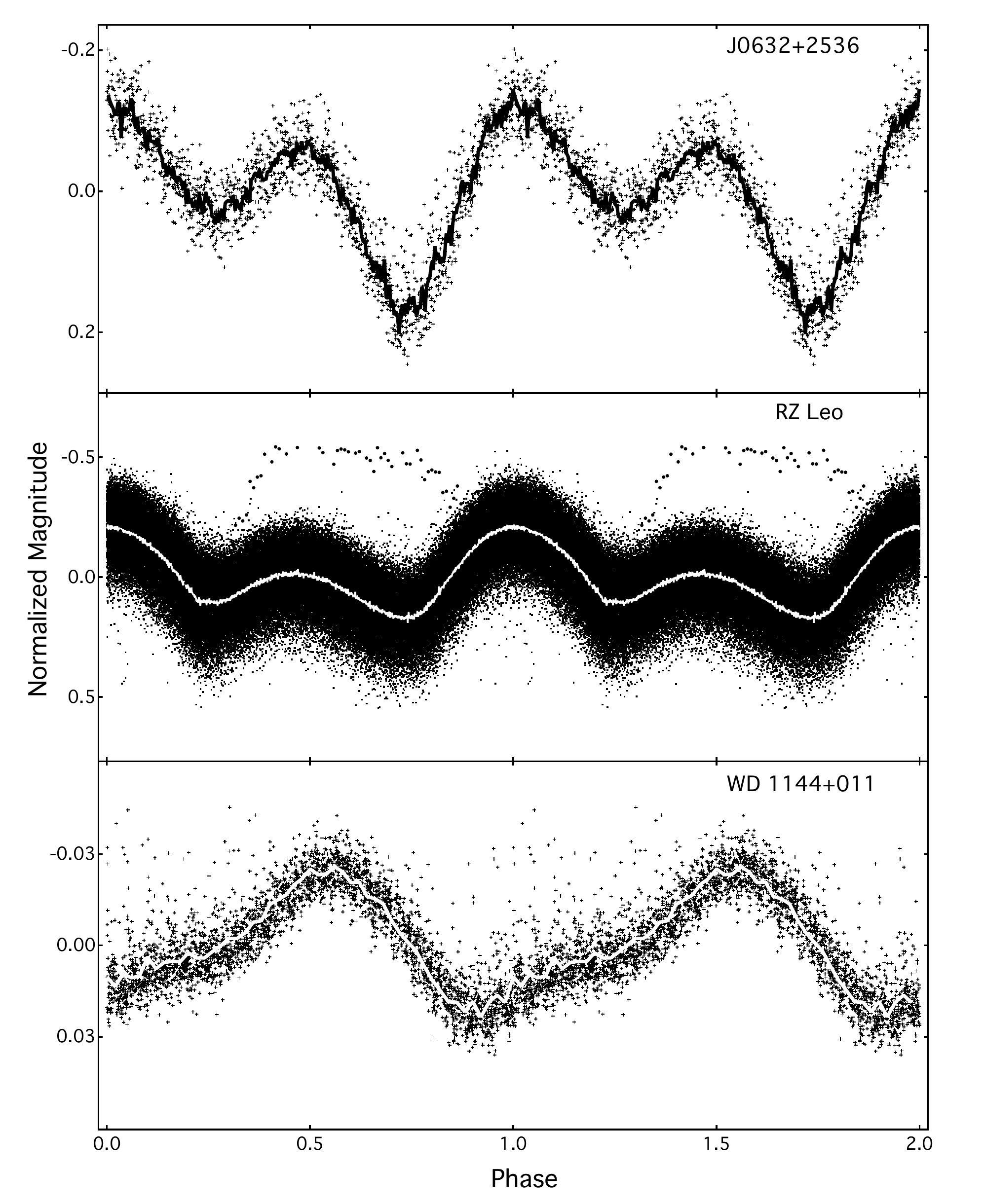}
\caption{\small{The phased and the binned light curves are plotted by the scatter crosses and solid lines, respectively. The brightening transient event of RZ Leo denoted by the dots is shown in the middle panel. The phases are arbitrary.}} \label{Figure 5}
\end{figure}

\begin{figure}
\centering
\includegraphics[width=16.0cm]{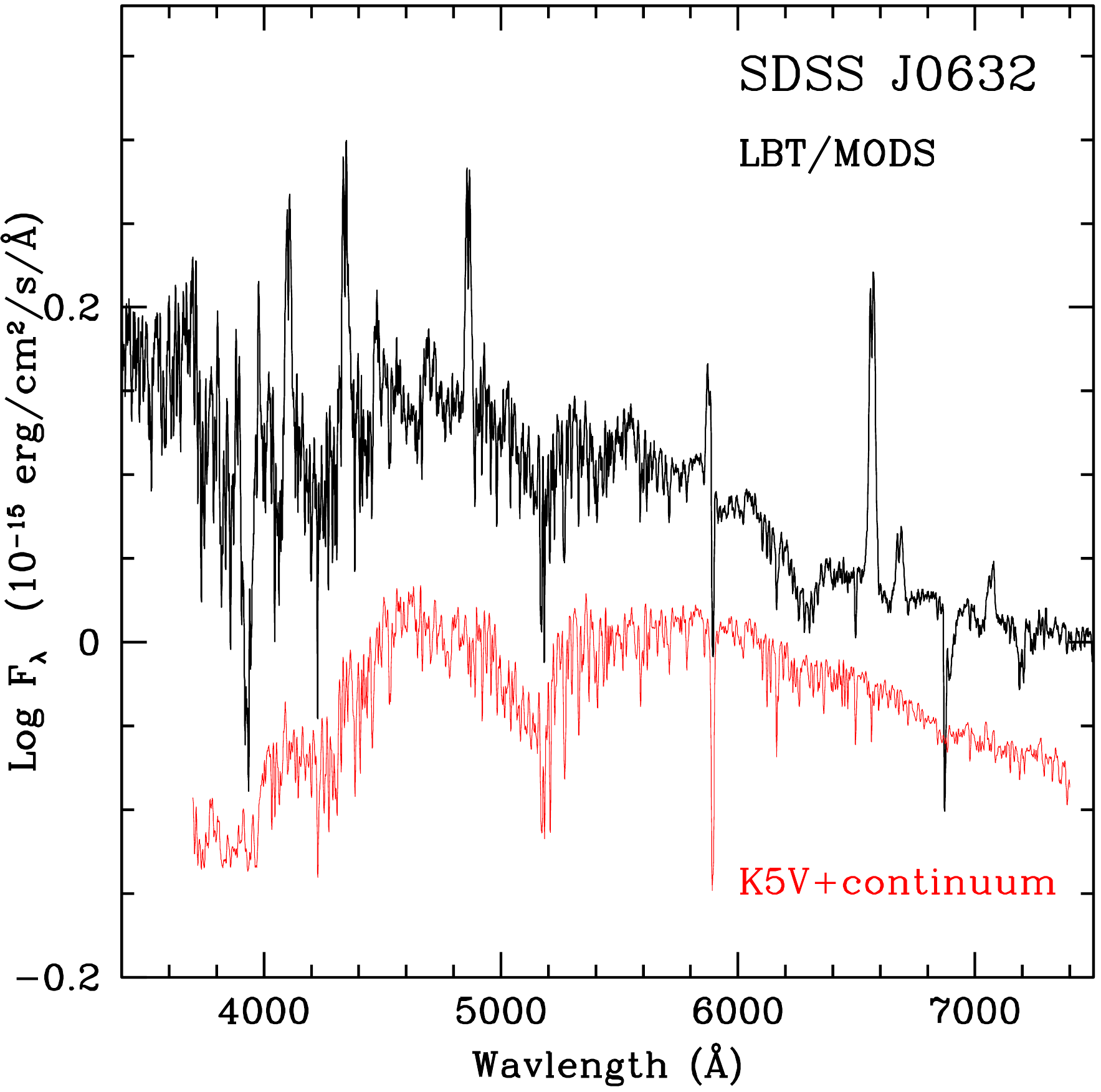}
\caption{\small{The optical spectra of J0632+2536 covering 3400\,-\,7400\,\AA\ taken by the LBT/MODS. The spectrum in red below is a template K5V spectrum by adding a sloping linear continuum. The Balmer series and the He II are doubled emission lines.}} \label{Figure 6}
\end{figure}

\begin{figure}
\centering
\includegraphics[width=16.0cm]{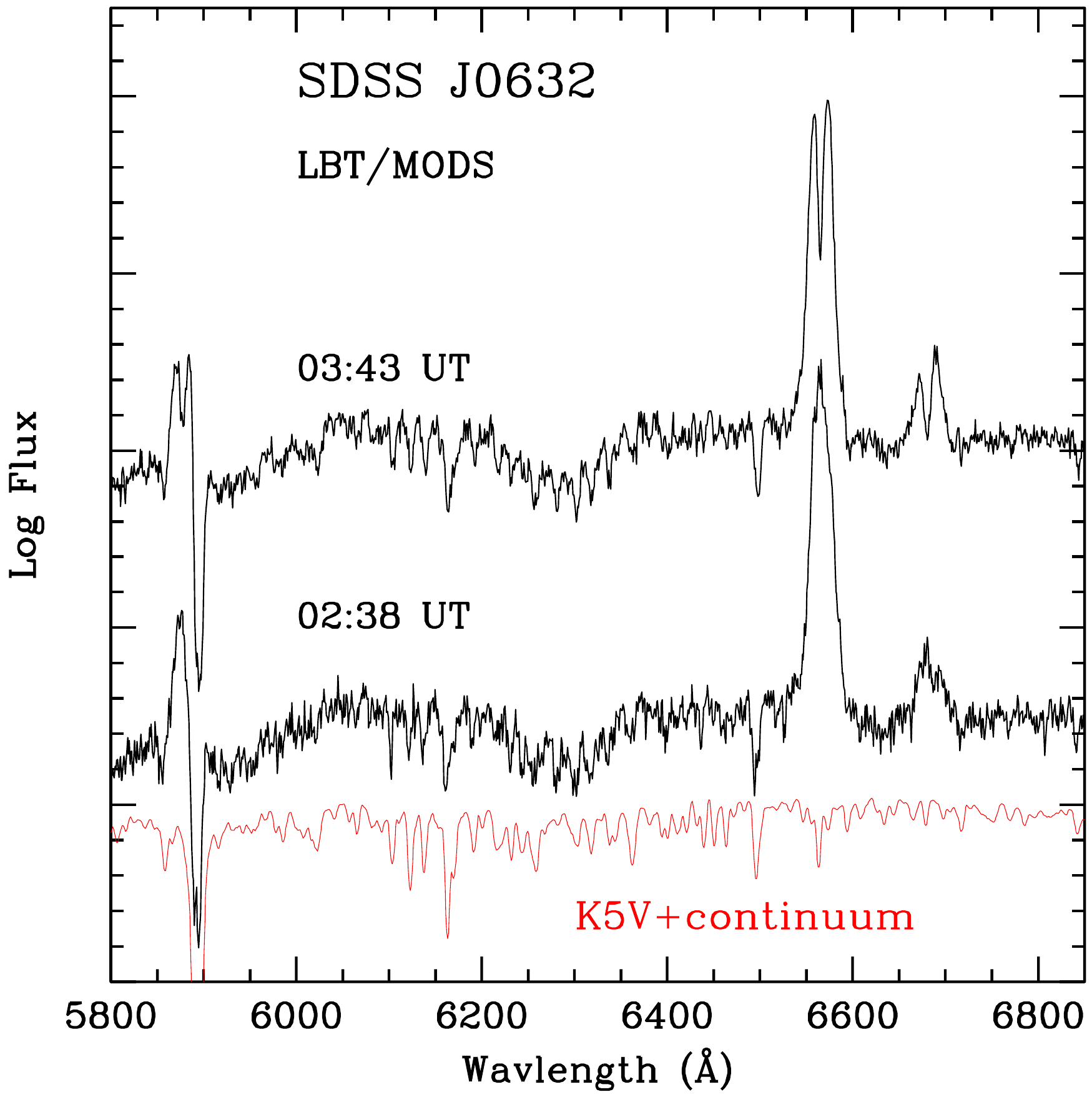}
\caption{\small{Two red LBT spectra of J0632+2536 covering 5800\,-\,6850\,\AA\  taken at 03:43 UT (top) and 02:38 UT (middle) show the doubled and single emission lines, respectively. The spectrum in red below is the same as Figure 6.} The unit of Log Flux is the same as that in Figure 6.} \label{Figure 7}
\end{figure}

\begin{figure}
\centering
\includegraphics[width=16.0cm]{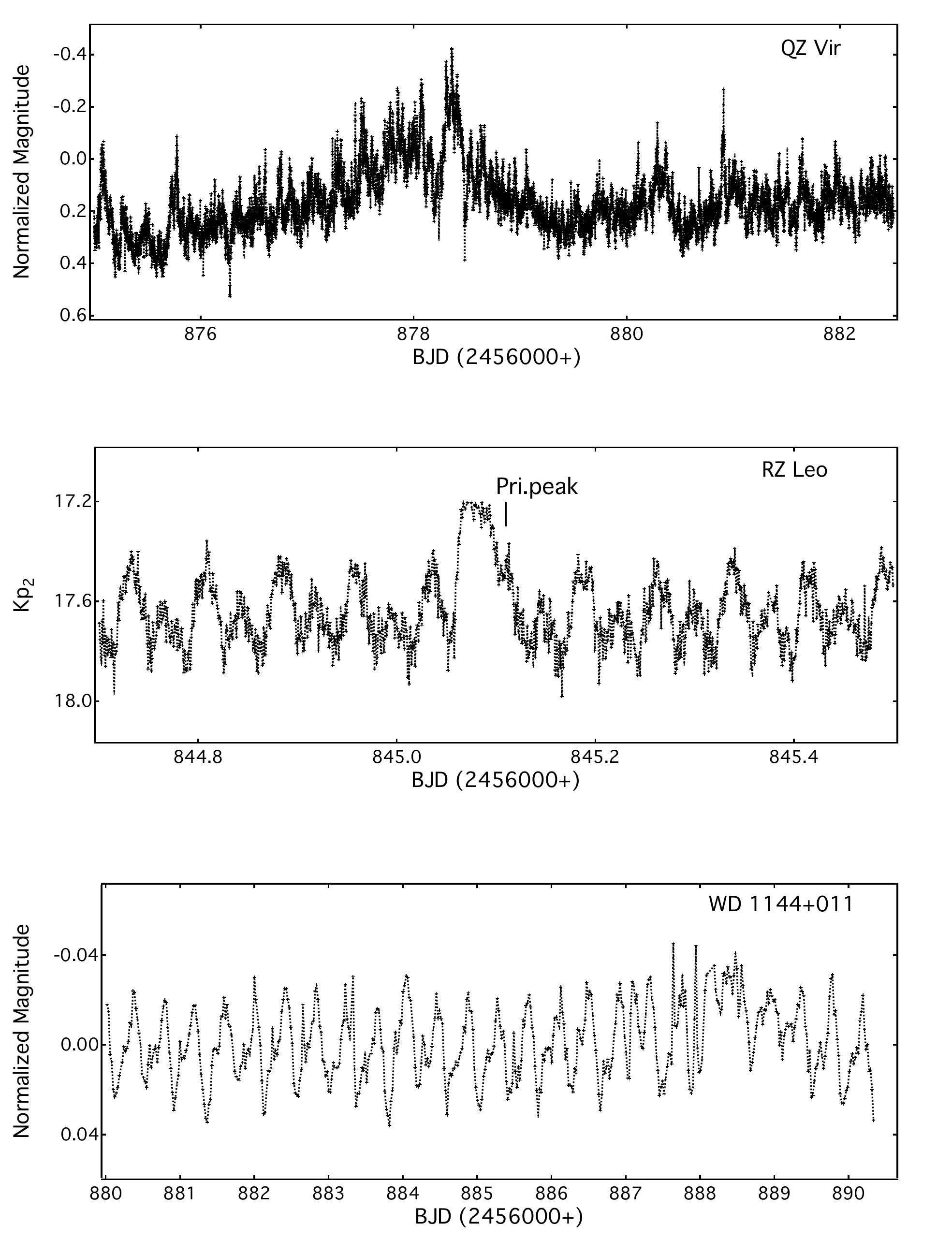}
\caption{\small{The zoomed-in light curves of the three transient events, which are marked by the dash rectangles in Figure 2. The small rebrightening event marked by the vertical line shown in the middle panel is regarded as a primary hump.}} \label{Figure 8}
\end{figure}

\begin{figure}
\centering
\includegraphics[width=15.0cm]{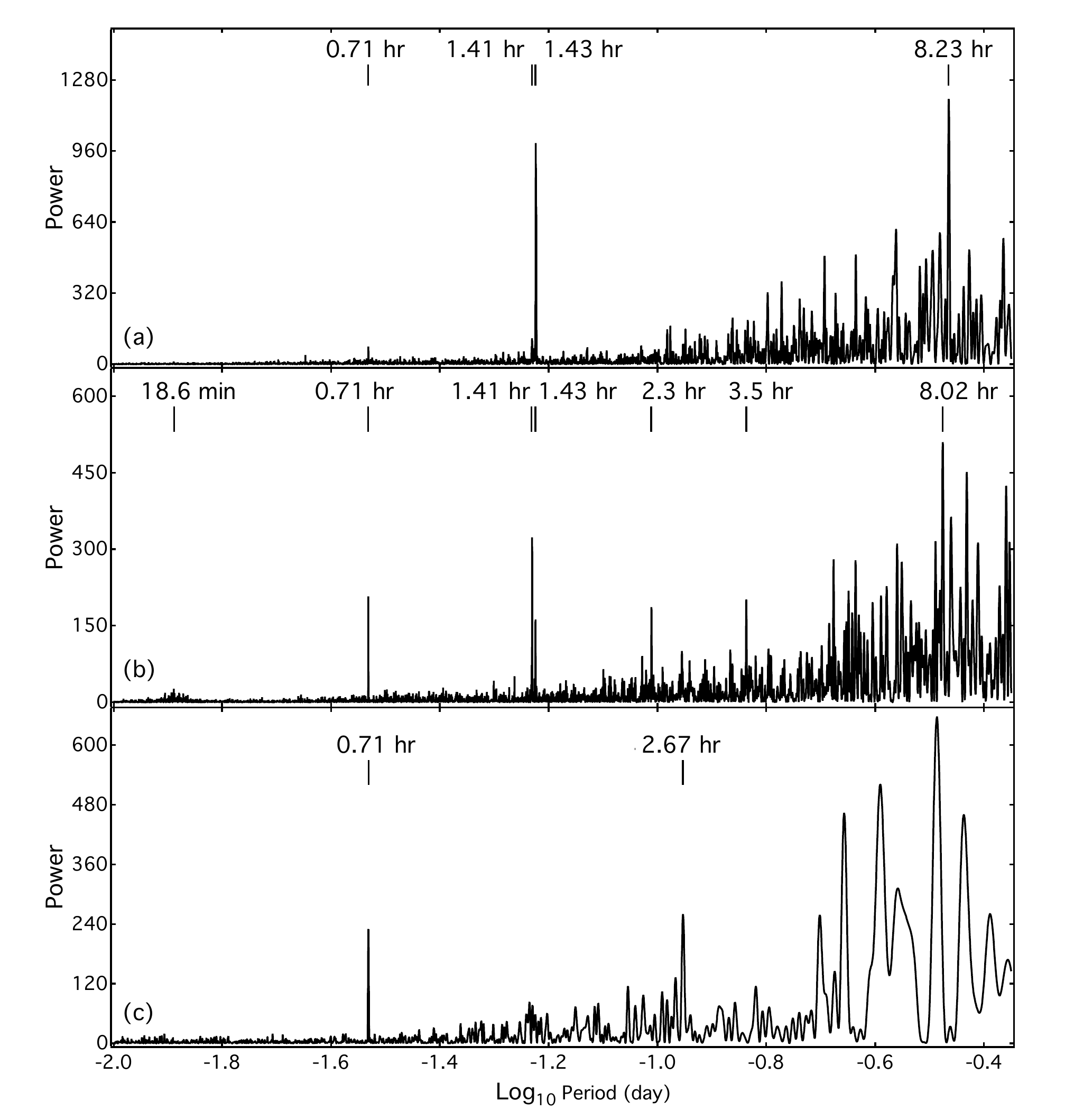}
\caption{\small{Periodograms of QZ Vir for the three light curves in different data ranges. (a) panel is for the assumed ending part of decay before the data gap of the K2-C1. (b) panel is for the quiescent light curve after the data gap. (c) panel is for the flare-like transient event shown in the top panel of Figure 8.}} \label{Figure 9}
\end{figure}

\begin{figure}
\centering
\includegraphics[width=12.0cm]{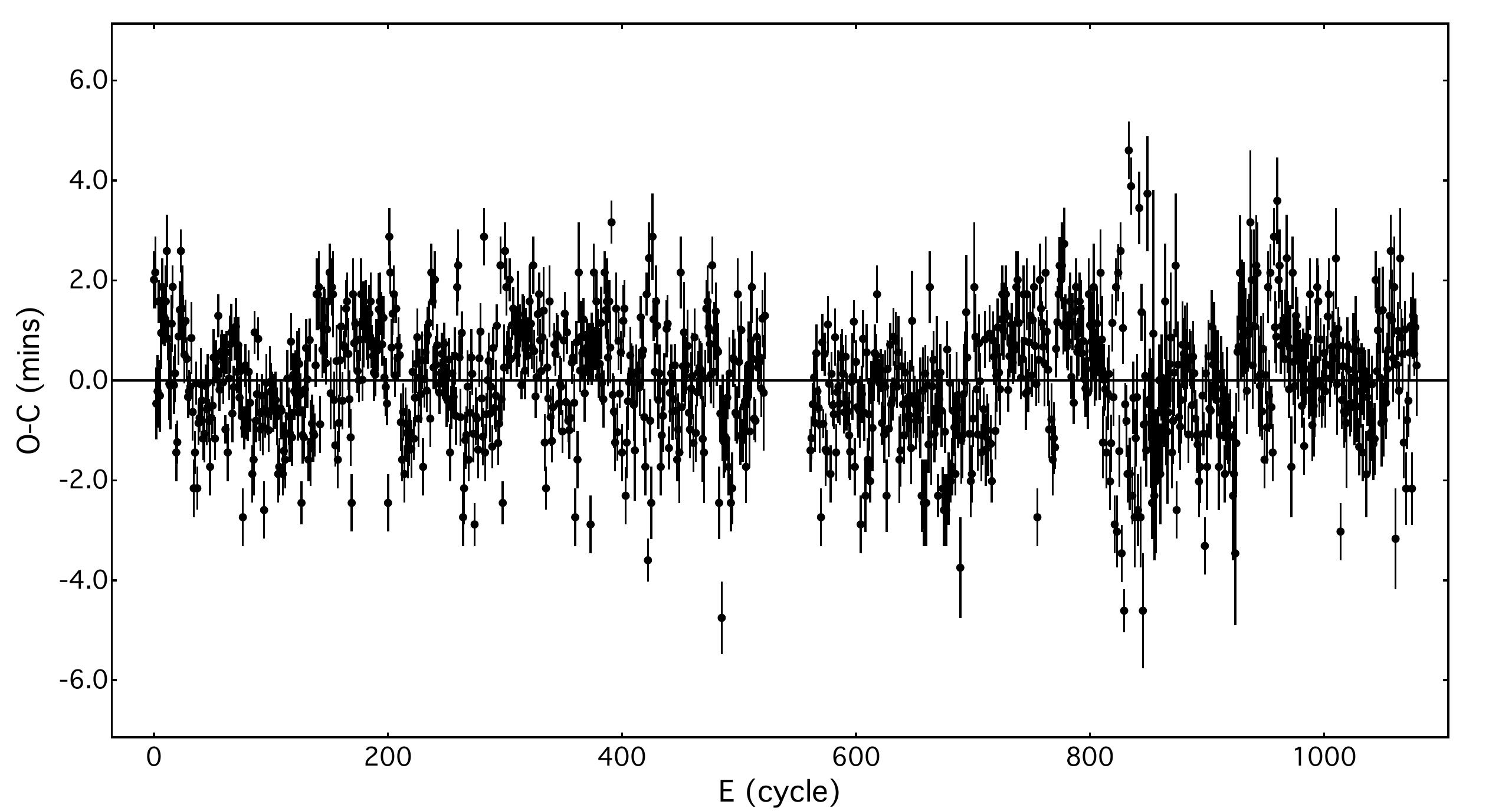}
\caption{\small{The O-C diagram for the maximum times of the primary humps of RZ Leo based on the improved ephemeris in Equation (2).}} \label{Figure 10}
\end{figure}

\begin{figure}
\centering
\includegraphics[width=15.0cm]{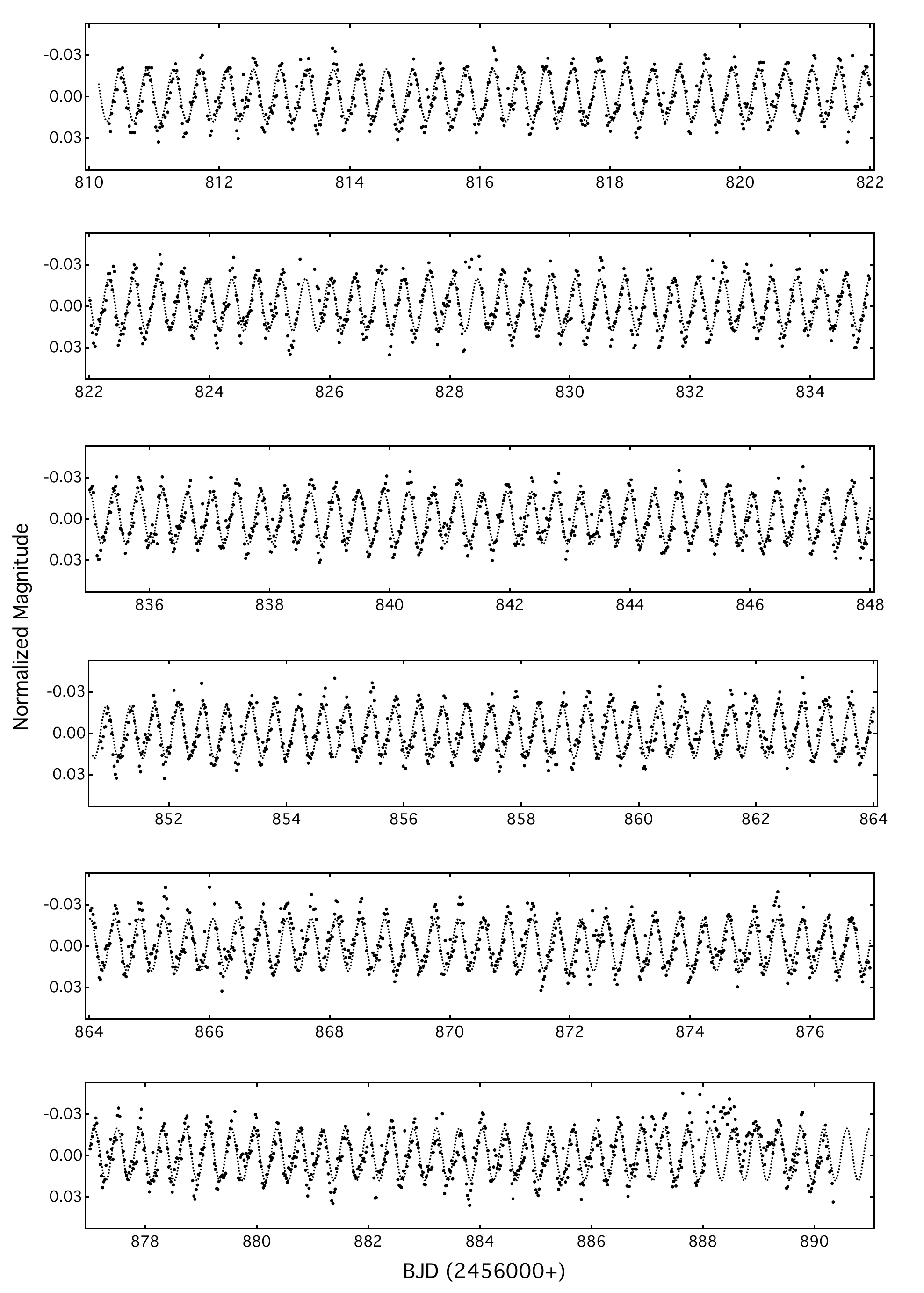}
\caption{\small{The best fitting sine curve and the normalized light curve of WD\,1144+011, which is shown in the bottom panel of Figure 2.}} \label{Figure 11}
\end{figure}

\begin{figure}
\centering
\includegraphics[width=15.0cm]{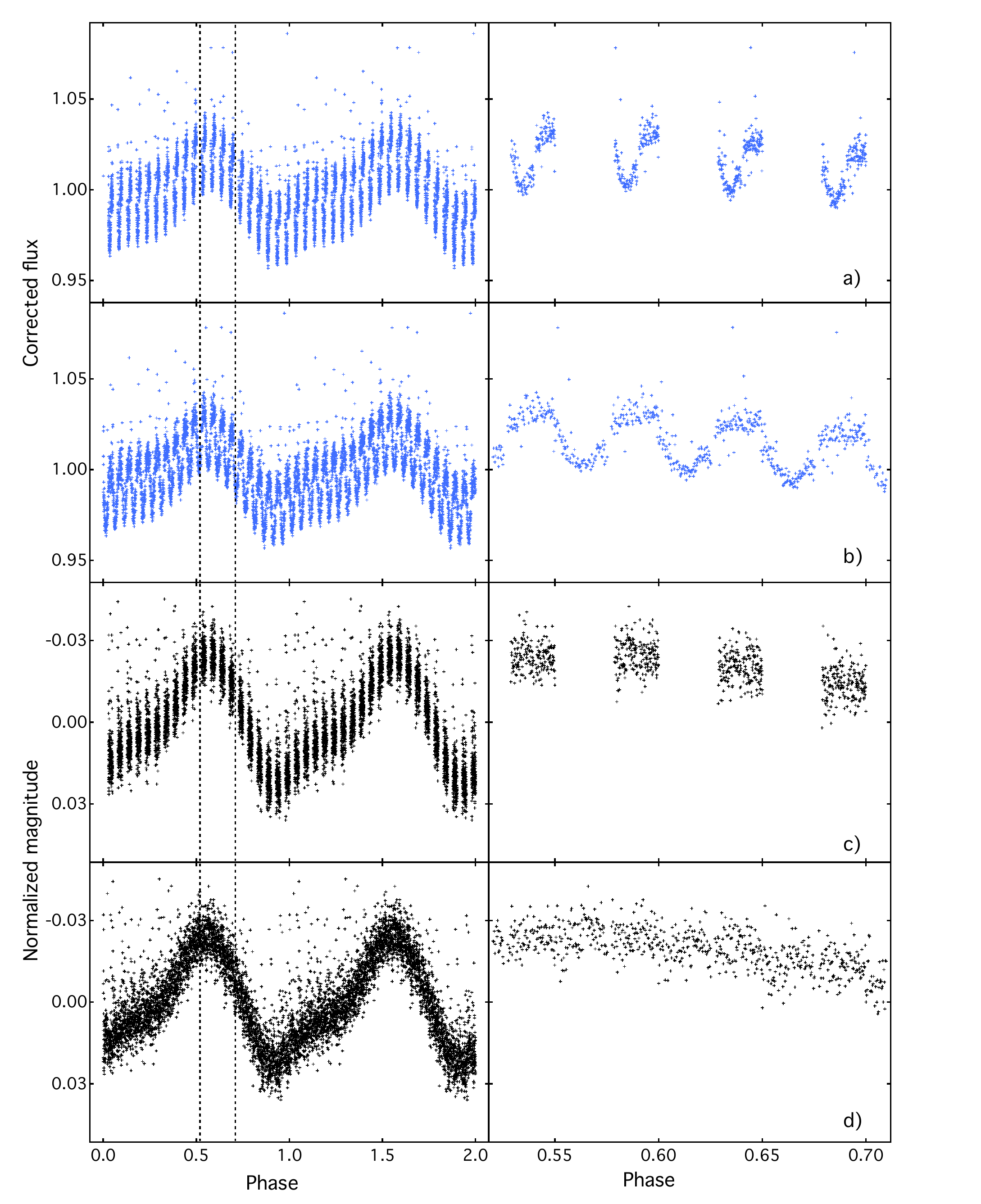}
\caption{\small{The blue phased light curves of WD\,1144+011 derived from the SFF corrected data are shown in the (a) and (b) rows, which correspond to folding periods of 0.408680655\,day and 0.408741\,day, respectively. The (c) and (d) rows present the phased light curves derived from our deduced K2 data based on the same folding periods as before. The light curves between the two dash lines in the four left panels are zoomed in their corresponding right-side panels.}} \label{Figure 12}
\end{figure}

\begin{figure}
\centering
\includegraphics[width=16.0cm]{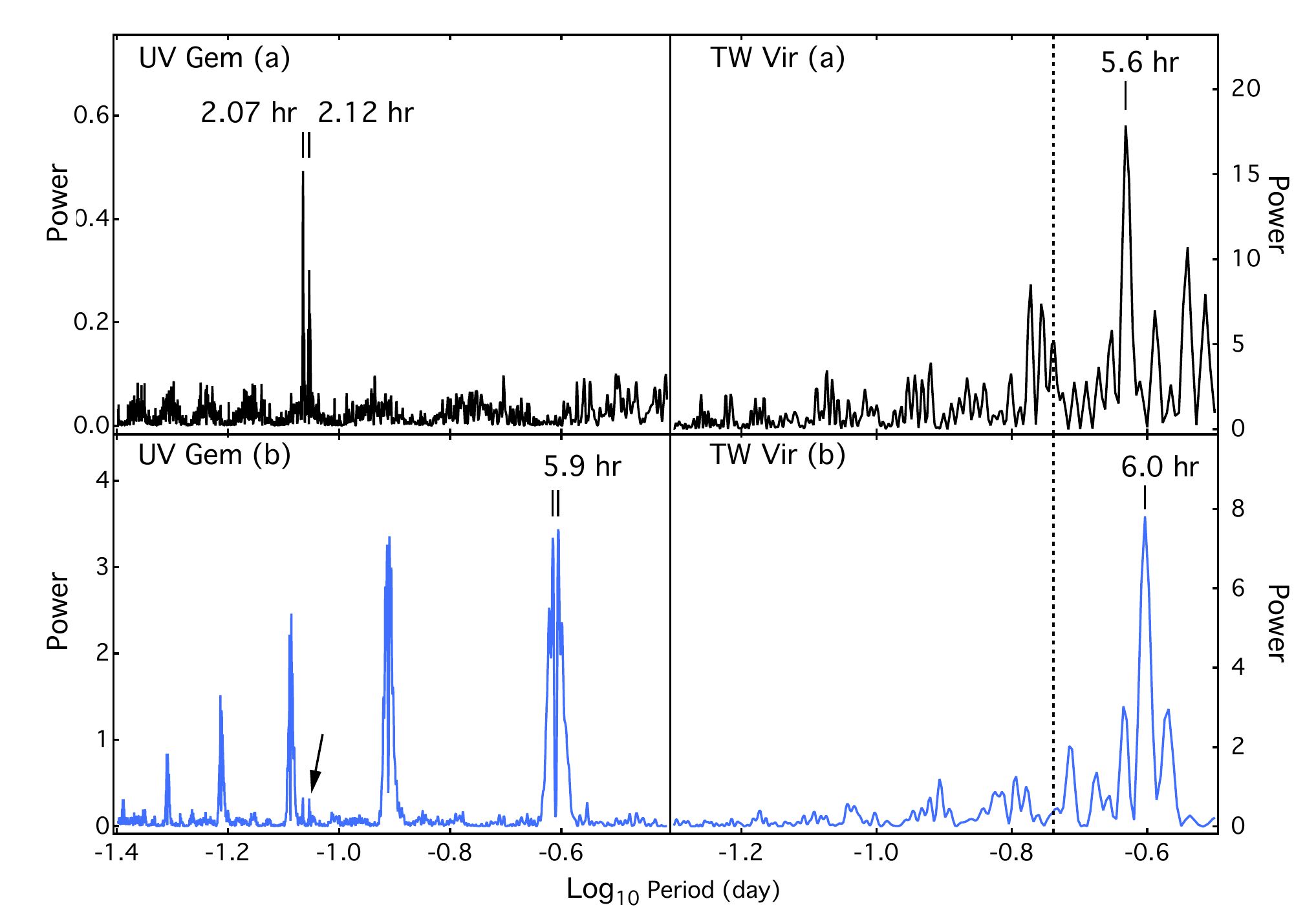}
\caption{\small{The periodograms for our deduced data and the SFF corrected data shown in the top and bottom panels are denoted in the blue and black lines, respectively, which are corresponding to their K2 light curves shown in Figure 1 and 3. Left panels: The periodograms of UV Gem for the whole K2 light curve shown in Figure 1. The two short vertical lines in the UV Gem (a) panel mark the plausible orbital period of UV Gem, which is also pointed out by the arrow in the UV Gem (b) panel. The 5.9 hr pseudo period is also marked in the UV Gem (b) panel. Right panels: The periodograms of TW Vir for the plateau of the superoutburst between the two dash lines shown in the bottom panel of Figure 2. The vertical dash lines in the right two panels refer to the orbital period 4.38\,hr of TW Vir derived by \citet{sha83}. The periods of 5.6\,hr and 6.0\,hr marked in the TW Vir (a) and (b) panels are the pseudo periods.}} \label{Figure 13}
\end{figure}

\begin{figure}
\centering
\includegraphics[width=15.0cm]{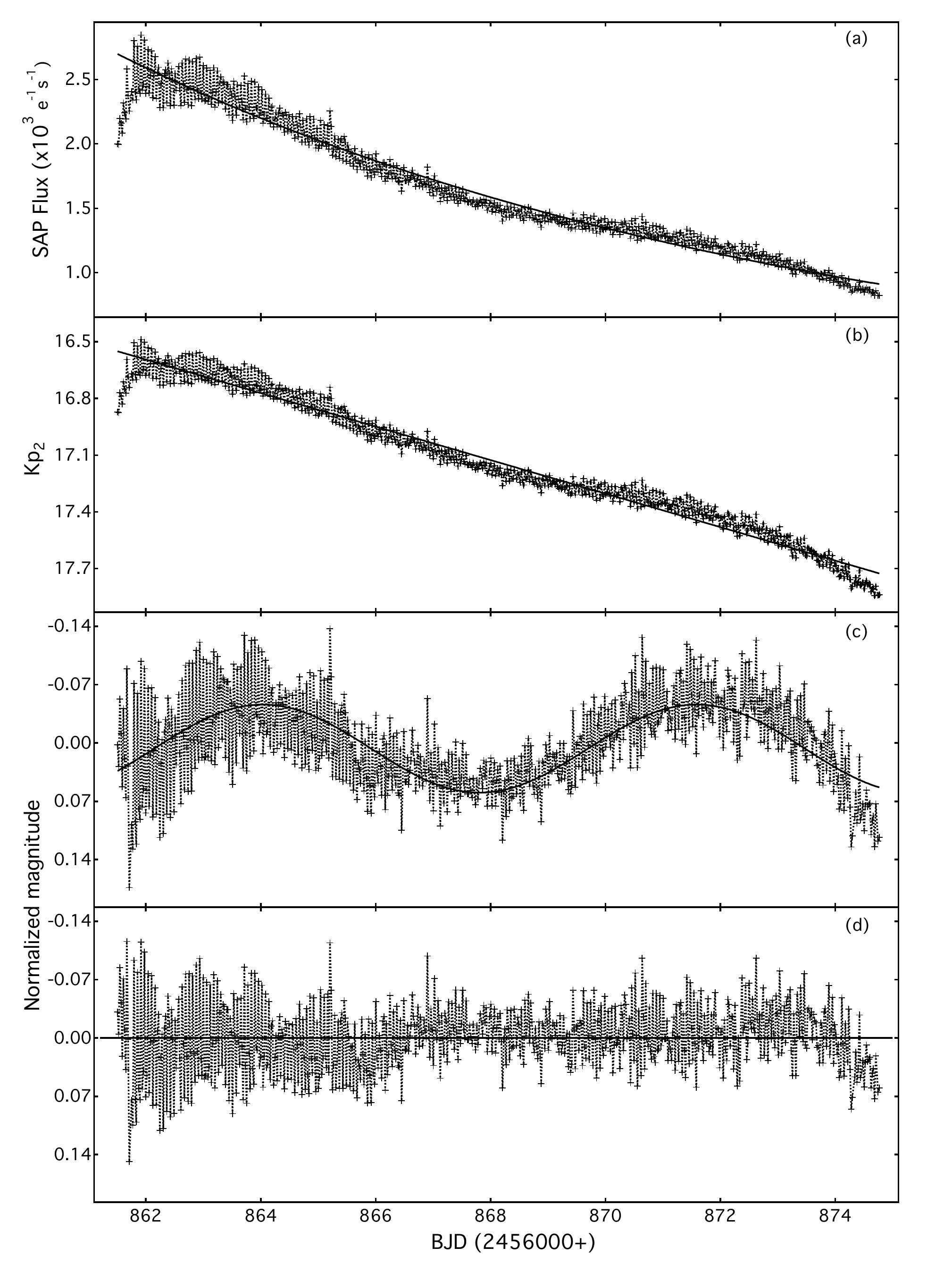}
\caption{\small{The zoomed-in plateaus of the super outburst of CSS1112+0028. The solid lines denote the best-fitting curves. From up to down, the (a) and (b) panels show the light curves in SAP Flux and Kp$_{2}$, respectively. The first normalized plateaus in magnitude by subtracting the linear line is plotted in the (c) panel. The second normalized plateaus after removing the sine variation is basically flat, which is also plotted in the (d) panel.}} \label{Figure 14}
\end{figure}

\clearpage

\end{document}